\documentclass[sigconf]{acmart}
\AtBeginDocument{%
  \providecommand\BibTeX{{%
    \normalfont B\kern-0.5em{\scshape i\kern-0.25em b}\kern-0.8em\TeX}}}

\copyrightyear{2024}
\acmYear{2024}
\setcopyright{rightsretained}
\acmConference[CHI '24]{Proceedings of the CHI Conference on Human Factors in Computing Systems}{May 11--16, 2024}{Honolulu, HI, USA}
\acmBooktitle{Proceedings of the CHI Conference on Human Factors in Computing Systems (CHI '24), May 11--16, 2024, Honolulu, HI, USA}
\acmDOI{10.1145/3613904.3641996}
\acmISBN{979-8-4007-0330-0/24/05}

\newcommand{\viewer}{viewer}
\usepackage{censor} %

\StopCensoring

\usepackage{longtable}
\usepackage{listings}
\usepackage{graphicx}
\usepackage{caption}
\usepackage{booktabs}

\lstdefinelanguage{yaml}{
    basicstyle=\ttfamily\footnotesize,
  columns=fullflexible,
  breaklines=true,
  breakatwhitespace=true,
  tabsize=2,
  literate={---}{{\textcolor{red}{}---}}{1},
}

\begin{document}

\title{Umwelt: Accessible Structured Editing of Multimodal Data Representations}

\author{Jonathan Zong}
\email{jzong@mit.edu}
\orcid{0000-0003-4811-4624}
\affiliation{%
  \institution{Massachusetts Institute of Technology}
  \streetaddress{32 Vassar St}
  \city{Cambridge}
  \state{Massachusetts}
  \country{USA}
  \postcode{02139}
}

\author{Isabella Pedraza Pineros}
\email{ipedraza@mit.edu}
\orcid{0009-0002-3269-7618}
\affiliation{%
  \institution{Massachusetts Institute of Technology}
  \streetaddress{32 Vassar St}
  \city{Cambridge}
  \state{Massachusetts}
  \country{USA}
  \postcode{02139}
}

\author{Mengzhu (Katie) Chen}
\email{mzc219@mit.edu}
\orcid{0009-0001-6404-7647}
\affiliation{%
  \institution{Massachusetts Institute of Technology}
  \streetaddress{32 Vassar St}
  \city{Cambridge}
  \state{Massachusetts}
  \country{USA}
  \postcode{02139}
}

\author{Daniel Hajas}
\email{d.hajas@ucl.ac.uk}
\orcid{0000-0002-2811-1197}

\affiliation{%
  \institution{Global Disability Innovation Hub}
  \streetaddress{1 Pool Street}
  \city{Stratford}
  \country{UK}
  \postcode{E20 2AF}
}

\author{Arvind Satyanarayan}
\email{arvindsatya@mit.edu}
\orcid{0000-0001-5564-635X}

\affiliation{%
  \institution{Massachusetts Institute of Technology}
  \streetaddress{32 Vassar St}
  \city{Cambridge}
  \state{Massachusetts}
  \country{USA}
  \postcode{02139}
}

\renewcommand{\shortauthors}{Zong et al.}

\begin{abstract}
We present Umwelt, an authoring environment for interactive multimodal data representations.
In contrast to prior approaches, which center the visual modality, Umwelt treats visualization, sonification, and textual description as coequal representations: they are all derived from a shared abstract data model, such that no modality is prioritized over the others.
To simplify specification, Umwelt evaluates a set of heuristics to generate default multimodal representations that express a dataset's functional relationships.
To support smoothly moving between representations, Umwelt maintains a shared query predicated that is reified across all modalities\,---\,for instance, navigating the textual description also highlights the visualization and filters the sonification.
In a study with 5 blind / low-vision expert users, we found that Umwelt's multimodal representations afforded complementary overview and detailed perspectives on a dataset, allowing participants to fluidly shift between task- and representation-oriented ways of thinking.
\end{abstract}

\ccsdesc[500]{Human-centered computing~Visualization toolkits}
\ccsdesc[500]{Human-centered computing~Accessibility systems and tools}

\keywords{multimodal data representation, visualization, textual description, sonification, accessibility}

\begin{teaserfigure}
    \centering{
  \includegraphics[width=0.9\textwidth]{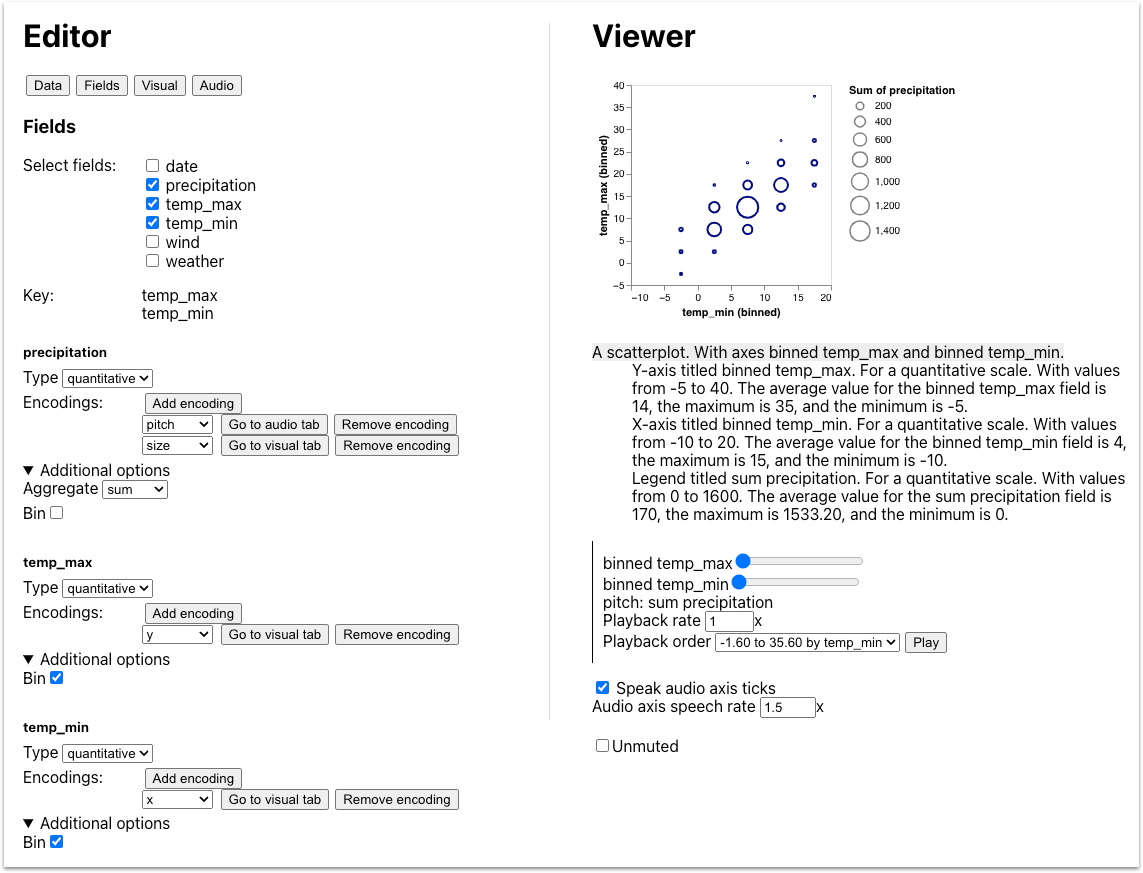}
  }
  \caption{A multimodal data representation designed in the Umwelt interface. Users specify fields and encodings in the accessible structured editor (left), which render into a visualization, textual structure, and sonification in the viewer (right).}
  \Description{
    A screenshot of the Umwelt interface with a two-column layout. The left column is the editor, featuring four tabs: `Data', `Field', `Visual', and `Audio'. The displayed content is from the `Field' tab, showing encodings for selected fields: `precipitation', `temp_max', and `temp_min' from the seattle-weather dataset. In the right column, the viewer displays a scatter plot, a textual description in the Olli tree structure, and audio controls. The scatter plot represents the sum of precipitation on the x and y axes, binned by temp\_min and temp\_max. The textual description has two visible levels: a summary of the scatter plot and summaries for the x-axis, y-axis, and legend. Audio controls include sliders for binned temp\_min and temp\_max, pitch control for precipitation sum, a configurable `Playback rate' currently set to 1, a `Playback order' dropdown, and a `Play' button. The `Speak audio axis ticks' checkbox is checked, and `Audio axis speech rate' is set to 1.5, adjustable via numerical input. At the bottom of the audio controls, there is an `Unmuted' checkbox.
    }
  \label{fig:teaser}
\end{teaserfigure}

\received{20 February 2007}
\received[revised]{12 March 2009}
\received[accepted]{5 June 2009}

\maketitle

\section{Introduction}

For blind and low-vision (BLV) people to be equal participants in interactive data analysis, they must be able to not only consume data representations created by others, but also create their own custom representations by rapidly prototyping and examining alternative designs~\cite{godfrey_advice_2015, hurst_empowering_2011}.
Critically, to have full agency over this process, BLV people must be able to independently author and understand data representations without relying on sighted assistance~\cite{godfrey_advice_2015}. 
In pursuit of these goals, accessibility research has begun to investigate multimodal data representation\,---\,that is, not only visualization but also textual description, sonification, and other modalities\,---\,with initial research results suggesting that the complementary use of multiple modalities can effectively facilitate analysis.
For instance, when both sonification and textual description are available, a screen reader user can get a high-level overview from sonification and use it to contextualize their detailed textual exploration \cite{fan_accessibility_2023}, and structured textual description helps low-vision magnifier users understand a scatterplot even when data points are visually occluded \cite{zong_rich_2022}.
Each modality structures information with different spatial and temporal trade-offs, so they often afford different tasks, comparisons, and navigation strategies.
Because of these modality differences, a screen reader user might want to author multiple representations to accomplish different goals, and easily switch between representations to develop a more holistic understanding of the data.

Unfortunately, existing tools for creating multimodal data representations center the visual modality: they assume the existence of a visualization that can \emph{then} be converted into an accessible representation. 
For instance, Chart Reader \cite{thompson_chart_2023} and VoxLens \cite{sharif_voxlens_2022} derive textual and sonified representations from an input specification of a visualization, while the SAS Graphics Accelerator \cite{noauthor_sas_2017} and Highcharts Sonification Studio \cite{cantrell_highcharts_2021} provide editors for non-visual representations that require users to first specify a visualization.
This ordering imposes limitations on both the authoring process and the expressivity of the output representations. 
In particular, it is challenging for a BLV person to independently create and interpret non-visual data representations unless they can first generate a corresponding visual chart\,---\,a problem that pervades existing statistical software \cite{godfrey_advice_2015}.
Moreover, a visualization-centric authoring process imposes an undue emphasis on replicating visual affordances non-visually by directly re-mapping encodings, instead of considering the distinct affordances of non-visual modalities.
As a result, this approach constrains the set of output representations that systems consider\,---\,for instance, Chart Reader and VoxLens restrict their support to a limited subset of chart forms that are straightforwardly amenable to sonification (e.g. bar charts, line charts) while sonifications based on other chart forms (e.g. scatterplots), or that diverge from the original chart's visual encoding (e.g. because they involve data transformations or interactions not specified in the visualization) remain underexplored.

In this paper, we present Umwelt%
    \footnote{The system is named after the concept of \textit{umwelt} in Jakob von Uexküll's semiotic theories \cite{uexkull_foray_2010}. An organism's \textit{umwelt} is the perceptual world produced by its subjective sense experience. Sense faculties vary across species; for example, bats hear ultrasound, and birds sense magnetic fields.
    As science journalist Ed Yong and disability activist Alice Wong note in an interview \cite{alice_wong_what_2022}, the idea of \textit{umwelt} does not support the notion that there is a normative sensory apparatus, either throughout nature or within the human species. Instead, it encourages us to equally value different subjective sense experiences.}%
: an authoring environment for multimodal data representations designed to de-center the visual modality.
A screen reader user can use Umwelt's \textit{structured editor} to specify data representations that include visualization, structured textual description, and sonification.
Instead of using a visual specification to generate non-visual representations, Umwelt derives each modality from a shared abstract data model.
As a result, users can create these representations in any order and/or specify only a subset of the three modalities as desired.
Moreover, via different sections in the editor, a user can switch between editing all modalities simultaneously, or making fine-grained edits to a particular modality.
To help users manage the upfront complexity of authoring a multimodal representation, the editor evaluates set of heuristics to generate default representations that express the dataset's functional relationships, and that a user can freely modify.
For example, a stocks dataset with a field \texttt{price} that depends on independent variables \texttt{date} and \texttt{symbol} will result in default representations that afford easily looking up the \texttt{price} for a given tuple of \texttt{date} and \texttt{symbol} --- a multi-series line chart, a textual structure that can group by \texttt{symbol} and \texttt{date}, and a sonification that plays back the \texttt{price} for each \texttt{date}, by  \texttt{symbol} (\autoref{fig:gallery}A).

The editor's state is rendered in Umwelt's \textit{\viewer} as independent visual, textual, and sonification views that are interactively linked together.
These interactions help maintain a shared context across modalities\,---\,for instance, navigating the text structure also highlights the corresponding data visually and filters the sonification domain to only play the selected values\,---\,and encourage users to think of the modalities as complementary views into the data.
Keyboard shortcuts help a screen reader user quickly move back and forth between the editor and \viewer{}, enabling a tight non-visual feedback loop for confirming the results of edits during prototyping.
The editor state is backed by an internal declarative specification (\autoref{fig:spec-plus-editor}).
This specification language describes an expressive space of multimodal representations in the \viewer{}. For example, Umwelt extends Vega-Lite's concept of view composition to express multi-view textual structures and sonifications.

We designed Umwelt through an iterative co-design process involving co-author \censor{Hajas}, who is a blind researcher with relevant expertise.
We evaluate our contribution with multiple evaluation methods, following best practices \cite{ren_reflecting_2018}.
Through an example gallery, we demonstrate that Umwelt's abstractions can express multimodal representations that span a variety of dataset semantics, data types, and view compositions.
We also conduct a study involving 5 expert BLV screen reader users to understand how the editor and \viewer{} help users conceptualize, author, and explore multimodal data representations. 
Our findings surface rich themes about how screen reader users approach multimodal data analysis.
We found that participants relied on complementary representations to move between overview and detail, and to manage cognitive and sensory load.
Interactive synchronization and runtime customizations enabled participants to access the data by reconfiguring and switching representations to use the one that best suited their immediate needs.
Participants also envisioned multimodal representations playing a role in facilitating communication between people who rely on different senses.
We also found that the editor reduced challenges associated with specifying representations, and surfaced different ways of thinking about the relationship between specifications and users' goals.

\section{Related Work}

Our work is informed by existing approaches to multimodal data representations and systems for accessibly authoring non-visual data representations. In this section, we briefly survey this pertinent literature to better characterize Umwelt's contributions.

\subsection{Multimodal Data Representations}
\label{sec:rw-multimodal}

Researchers and practitioners have explored a variety of approaches to data representations beyond visualization.
Some systems have focused one one primary alternate modality\,---\,for example, Olli~\cite{blanco_olli_2022} explores how textual descriptions can be structured to provide varying levels of detail~\cite{zong_rich_2022}. 
A larger body of systems has explored how multiple non-visual modalities can be used in concert. 
For instance, Apple's VoiceOver Data Comprehension feature on iOS~\cite{davert_whats_2019} offers out-of-the-box support for making data accessible through verbal descriptions and sonification (or non-speech audio).
Similarly, research systems have explored methods for combining tactile graphics with voice \cite{baker_tactile_2014, baker_tactile_2016}, sonification with voice \cite{holloway_infosonics_2022}, haptics and sonification \cite{fan_slide-tone_2022}, and sonification and interactive question-answering.
Among such multimodal systems, Chart Reader~\cite{thompson_chart_2023} is a particularly apt point of comparison to our work because, like Umwelt, it incorporates best practices in visualization, structured textual description, and sonification into a single analysis interface.

While these systems make important and necessary contributions to accessible visualization, they share a common assumption: they begin with a visual artifact and attempt to retarget visual affordances to non-visual modalities.
For example, Olli, VoxLens, and Chart Reader all require a visualization specification as input to generate their non-visual representations.
As a consequence, these systems are often unable to express data representations that do not have an analogous visualization. 
Chart Reader, for instance, can only express sonifications that directly correspond to the specific typology of chart types it supports.
In contrast, Umwelt does not derive its non-visual representations from the visual specification.
Instead, its three modalities are treated as equal outputs, all derived from an abstract data model that is shared across modalities.
In \autoref{sec:examples}, we show examples of multimodal representations rendered in the Umwelt viewer that exceed the expressiveness of prior systems (e.g. because their audio specification diverges from the visual).

\subsection{Accessible Authoring Tools for Non-Visual Data Representations}

In contrast to tools that convert an existing artifact into another representation, researchers have also explored authoring toolkits for multimodal representations.
However, most existing toolkits correspond to a single non-visual modality.
For example, Highcharts Sonification Studio \cite{cantrell_highcharts_2021} is an authoring tool for producing charts with sonification, and SVGPlott \cite{engel_svgplott_2019} is an authoring tool for tactile charts.
Though these tools are designed to author non-visual data representations, they require a user to specify a visualization to convert into a non-visual form.
Consequently, they suffer from the same expressiveness issues discussed in \autoref{sec:rw-multimodal}.
For example, SAS Graphics Accelerator \cite{noauthor_sas_2017} includes an authoring workflow that makes charts accessible via sonification and textual description, yet does not support sonification for many chart types.

Because these authoring environments require users to specify visualizations, they impose additional demands on BLV users.
For instance, users must have a visual form in mind before creating a non-visual representation, and need an accessible way to verify the accuracy of their visual specification.
Umwelt addresses these concerns by allowing users to create representations in any order, and without requiring users to specify all three modalities.
Instead, representations are authored independently, reducing the need to conceptualize all outputs in terms of the visual modality.
For example, when a user loads a dataset, a textual structure will be generated describing the data in terms of its fields.
They can directly specify a sonification by assigning audio encodings, without needing to specify them in terms of visual concepts like the x- or y-axis.

Of existing authoring toolkits, the closest point of comparison is PSST \cite{potluri_psst_2022}, which enables BLV users to create multimodal representations of streaming data that include sonification, spoken description, and physical laser-cut artifacts.
Just as we propose with Umwelt, PSST does not require a visual specification.
However, PSST differs from Umwelt in terms of its level of abstraction; where Umwelt offers a higher-level workflow and abstractions, PSST exposes low-level abstractions such as event streams, handlers, and a dataflow graph.
This has consequences for how users conceptualize and author representations; in \ref{sec:qualitative-results}, we discuss how moving between field- and encoding-oriented specification in Umwelt's editor helped users reason about data in terms of both tasks and representations.

\section{Motivation: De-Centering the Visual Modality}
\label{sec:motivation}

In this section, we discuss how the overarching motivation of de-centering the visual modality in data analysis translates to concrete design goals for multimodal authoring systems.
We then walk through an example usage scenario to demonstrate how instantiating these design goals in Umwelt enables a user to conduct independent data exploration using multiple complementary modalities.

\subsection{Design Goals}
\label{sec:design-goals}

We designed Umwelt through an iterative co-design process led by co-authors \censor{Zong} and \censor{Hajas}. \censor{Hajas} is a blind researcher with relevant experience in designing accessible representations.
Over the course of about a year, we developed multiple prototypes of various interactive sonification and textual description techniques, accessible editor interfaces, and syntax prototypes of Umwelt's abstract model.
All co-authors discussed prototypes regularly over Zoom call and email, reflecting on their strengths and weaknesses and brainstorming directions for additional iteration.

Early in the design process, we identified several challenges in the design of an authoring environment for multimodal data representations that arose from the core motivating principle of de-centering the visual representation.
We synthesized these challenges into a set of design goals (DGs) that guided our iterative process and influenced team discussions where we reflected on candidate designs.
Sections \ref{sec:editor} and \ref{sec:viewer} will elaborate on how these design goals are addressed in the editor and \viewer, respectively.

\begin{enumerate}
    \item \textbf{Deferred commitment to a modality.} Authors often do not begin a rapid prototyping process with a concrete idea of their desired end state.
    As such, it is important for an authoring tool to offer the flexibility to easily try many candidate representations in an exploratory manner.
    In a multimodal system, this might involve freely editing different modalities in any order, or easily changing a field's mappings from one modality to another as they explore possible designs.    
    In existing systems, the non-visual modalities depend on a visual modality. This requires an author to \textit{prematurely commit} \cite{blackwell_cognitive_2001} to a visual representation before specifying other modalities.
    Our goal is instead to encourage deferred commitment.
    For example, an author should be able to specify non-visual modalities independently without first needing to create a visual specification.
    
    \item \textbf{Complementary use of modalities.} 
    An advantage of a multimodal system is that users are not required to rely on a single representation to meet all of their needs.
    Due to differences in how each modality conveys information, it is difficult to expect any single representation to act as a standalone replacement for another.
    Instead, a goal of our system is to encourage users to choose the modality that best suits their task at any given time, or use multiple modalities together to gain a more complete understanding.    
    
    \item \textbf{Common ground between mixed-ability collaborators.} In a system designed to de-center the visual representation, the visualization still serves an important purpose for collaboration between people who primarily use different senses. For instance, screen reader users are not a monolith; people who use both a screen reader and a magnifier, like many low-vision users do, benefit from referencing a visual chart alongside other representations. Additionally, an important possible use of data analysis is to communicate about it with other people in personal or professional settings. BLV data analysts may need to communicate their findings to audiences that include people with different levels of vision, or participate in discussions where sighted colleagues are referencing visual concepts.
    Consequently, we find it important that representations establish common ground for diverse collaborators.
\end{enumerate}

\subsection{Authoring Multimodal Representations Co-Equally: An Example Usage Scenario}
\label{sec:walkthrough}

\begin{figure*}
  \centering
  \includegraphics[width=\textwidth]{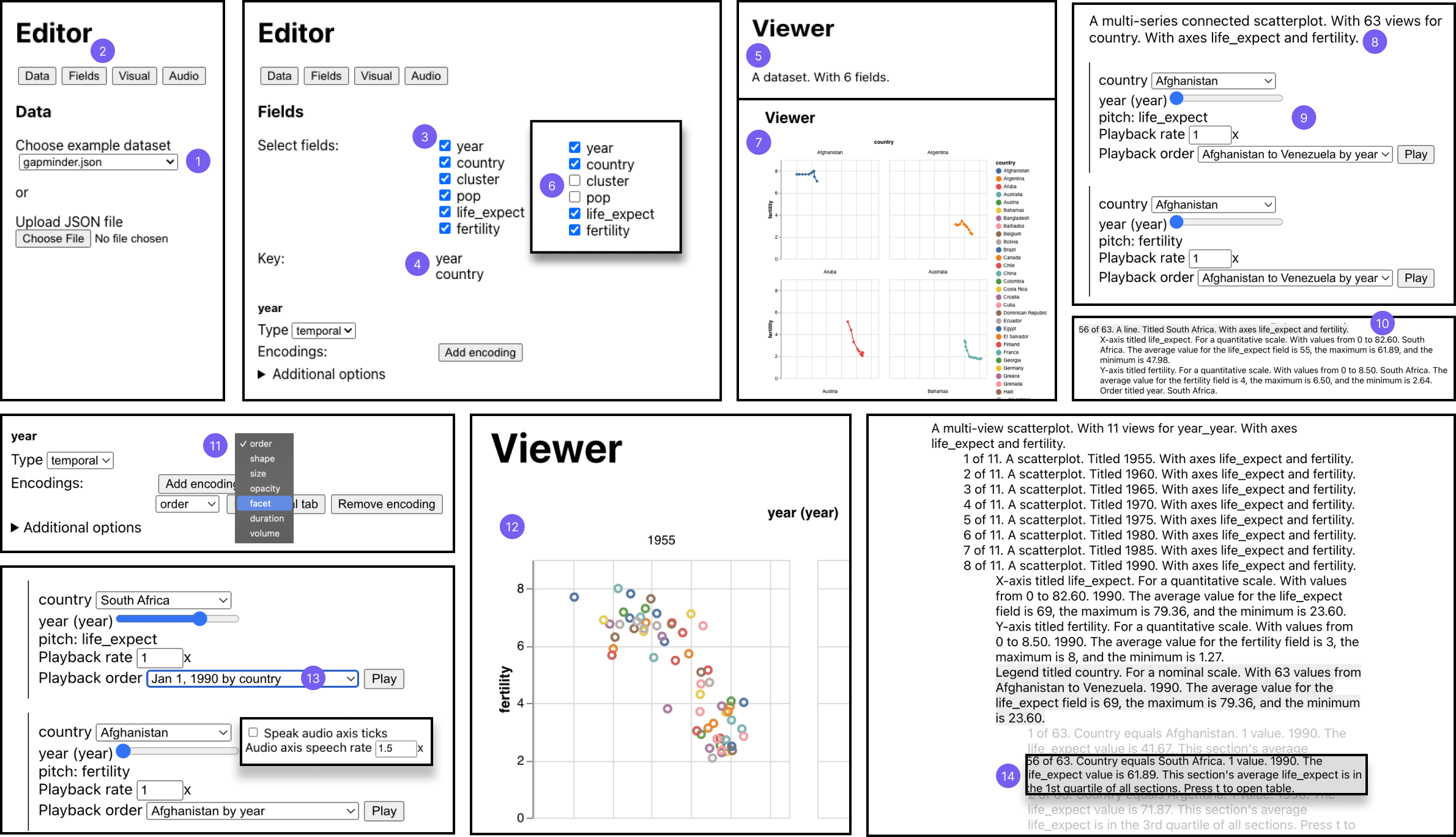}
  \caption{An analyst's workflow in Umwelt: 1--6 illustrate an analyst's process of creating an initial multimodal data representation (shown in 7-9). 10 shows their initial exploration, before (11) making edits and then (12--14) continuing their analysis .}
  \Description{
  A set of Umwelt walkthrough screenshots is organized into two sections. The top row demonstrates a user creating a data visualization, which covers loading data, customization of fields and encodings, and displaying results with an Olli tree and small-multiple connected scatter plot, along with two sets of audio controls. The bottom row showcases a user exploring the dataset using both audio controls and text descriptions. On the left, it shows the user altering encodings and playback order, while the right side depicts real-time updates to the visualization and corresponding textual descriptions of the Olli tree.
  }
  \label{fig:walkthrough}
\end{figure*}

To demonstrate the process of authoring and analyzing multimodal representations in Umwelt, we walk through a scenario in which a screen reader user named Lula explores Hans Rosling's well-known Gapminder dataset \cite{rosling_best_2006}.

\textbf{Field-driven default specification.}
When Lula loads the dataset (\autoref{fig:walkthrough}.1) into the editor in the \textsc{Data} tab, the \textsc{Fields} tab (\autoref{fig:walkthrough}.2) is populated with all of the dataset's fields.
Lula inspects the set of checkboxes labeled ``select fields'' to ascertain that all six fields are initially checked (\autoref{fig:walkthrough}.3) and thus are participating in the default multimodal representation. 
Reading the section below these checkboxes, Lula learns that Umwelt has inferred a composite key of \texttt{(year, country)} (\autoref{fig:walkthrough}.4).
Jumping over to the \viewer{} by pressing the `v' key on their keyboard, they observe that the initial representation is a textual structure that hierarchically groups and organizes the data for each field (\autoref{fig:walkthrough}.5).

Lula decides that they want to analyze the life expectancy vs. fertility rate of  countries over time, mirroring Rosling's original global health scatterplot \cite{rosling_best_2006}.
They jump back to the editor with the `e' key and tab through the checkboxes to keep only \texttt{year}, \texttt{country}, \texttt{life\_expect}, and \texttt{fertility} checked (\autoref{fig:walkthrough}.6).
Umwelt's infers a new multimodal representation by reasoning about the dataset's keys and the measure types of the selected fields (i.e., nominal, quantitative, etc.).
For the fields Lula has selected, as the data is keyed by \texttt{(year, country)}, Umwelt assumes that a typical reader will use these fields to lookup the value fields \texttt{(life\_expect, fertility)}.

Though there may be multiple ways to represent the same key-value semantics, the goal of Umwelt's heuristics is to provide an initial starting point rather than a single best representation.
Thus, for the selected fields, Umwelt produces a multimodal representation that includes a small multiple of connected scatterplots (\autoref{fig:walkthrough}.7), a textual structure (\autoref{fig:walkthrough}.8), and two sets of audio controls (\autoref{fig:walkthrough}.9), such that all modalities support this lookup operation via their modality-specific affordances.
The visualization facilitates the lookup by faceting the data into multiple views by the \texttt{country} field, and using the \texttt{year} to order a connected scatterplot in each view.
The sonification supports this same lookup by offering two sets of audio controls\,---\,each corresponding to an \textit{audio unit}, or a single audio track that plays a continuous tone with pitch corresponding to \texttt{life\_expect} or \texttt{fertility} respectively.
Both audio units allow Lula to select \texttt{(year, country)} tuples, either via sequential playback by pressing the play button, or via interactive selection by manipulating a dropdown and slider.
Finally, the textual structure facilitates the key-value lookup via its hierarchical structure.
The first level below the root allows Lula to choose a \texttt{country}, and the next level allows Lula to drill further down into \texttt{year}, \texttt{life\_expect}, or \texttt{fertility} (\autoref{fig:walkthrough}.10).
This hierarchy is generated based on the fields and key, and would exist even without a visual representation; however, because visual information is present, it's used to annotate the fields' descriptions in terms of x, y, and order encodings.
At the lowest levels of the tree, Lula receives summary information about the selected data, and
they can also press `t' to open a tabular view of the data.

\textbf{Analyzing data using multiple modalities.}
Lula starts by pressing `p' on their keyboard to play the sonification for \texttt{life\_expect}. 
Umwelt orients them via \textit{audio axis ticks}, which are spoken announcements of data values interleaved with the sonification to communicate playback progress.
The system announces the first \texttt{country}, Afghanistan, and as the sonification progresses, the system speaks the \texttt{year} value prior to the sonified tone for each 5 year interval.
After listening through a few \texttt{country} values, Lula understands how to interpret the sounds, and disables the  audio axis ticks feature  to more rapidly get a gist of the rest of the data.
They observe that \texttt{life\_expect} generally increases for most countries, but the min and max values can vary widely.
For a few countries that sound different from the rest, Lula pauses the sonification by pressing `p' on their keyboard again.
To determine which country they were listening to, Lula tabs to the set of audio controls for \texttt{life\_expect}, and inspects the \texttt{country} dropdown menu which reflects the current position in the paused playback.

Noticing that South Africa's \texttt{life\_expect} sonification peaks in the middle before dropping again, Lula jumps to the textual structure with the `o' key and navigates to the corresponding node by using the down arrow to move from the root level to the \texttt{country} level, and using the left and right arrow keys to find the node representing South Africa.
Descending a level to the x-axis, they read the exact average, min, and max summary values of \texttt{life\_expect} for South Africa\,---\,grounding the sonification they heard before in concrete numbers.
By navigating to a sibling branch of the textual hierarchy, Lula is instead able to step through each year to read the exactly value of \texttt{life\_expect} from \texttt{1955} to \texttt{2005}.
To remind themselves of the overall trend, they press `p'. 
Their position in the textual structure emits a query predicate that filters the sonification domain and highlights the corresponding data in the visualization.
Because their cursor focus remains on the textual description for South Africa over a particular set of years, the sonification only plays through this data subset.
Thus, they are able to identify that the peak they heard was for \texttt{1990}, when \texttt{life\_expect} was \texttt{61.89}. 

\textbf{Editing the representation design.}
Umwelt's heuristics for these four fields prioritize using \texttt{country} as the ``outermost'' key, or the key that is used first in the composite key lookup operation.
In other words, the multimodal representations afford looking up a specific \texttt{country} before exploring it by \texttt{year}.
Now that Lula has done that, they may want to explore the data other way: by picking a \texttt{year} and exploring it by \texttt{country}.

To do so, Lula jumps back to the \textsc{Fields} tab of the editor, and removes the \texttt{facet} encoding from the definition for \texttt{country}.
Umwelt updates the visualization to a single view containing a multi-line connected scatterplot (also known as a trace visualization \cite{robertson_effectiveness_2008}).
The textual descriptions update to reflect this change, though the hierarchical structure remains unchanged; the sonification, similarly, does not change.
Lula then updates the definition for \texttt{year}, switching the \texttt{order} encoding for a \texttt{facet} encoding instead (\autoref{fig:walkthrough}.11).
As a result, the visualization is faceted by year and each facet contains a scatterplot (\autoref{fig:walkthrough}.12) with one point per country.
The textual hierarchy now updates with \texttt{year} at the first level, and \texttt{country} nested underneath.
The sonification still does not change as its traversal ordering is based on the key, which has remained the same throughout.
Lula is able to verify these edits had the intended effect by quickly jumping back and forth between the \viewer{} and editor with the `e' and `v' hotkeys.

Lula then repeats their preferred analysis process of sonification overview followed by detailed textual exploration.
To listen to the \texttt{life\_expect} values for each \texttt{country} in a given \texttt{year}, they first select a \texttt{year} using the slider\,---\,they start with \texttt{1990} to see what else was happening in the world during their previously observed notable year.
Then, using the ``playback order'' dropdown, they select ``1990 by country'' (\autoref{fig:walkthrough}.13).
After turning ``speak audio axis ticks'' back on and pressing play, they hear the name of each \texttt{country} followed by a short tone corresponding to its \texttt{life\_expect} in \texttt{1990}.
This gives them a general sense of the variability of \texttt{life\_expect} values in \texttt{1990}.
Listening for South Africa, they have a sense of the relative position of that tone compared to the higher or lower tones that they've heard.
Returning to the textual description, they navigate to \texttt{1990} and drill down into the \texttt{country} legend. They navigate to South Africa and are reminded of its average \texttt{life\_expect}, and read that this value is in the 1st quartile of \texttt{life\_expect} values --- meaning that it is below the 25th percentile of values (\autoref{fig:walkthrough}.14).

\textbf{Summary.}
Using Umwelt, Lula was able to author multimodal data representations involving visualization, structured textual description, and sonification as part of a self-guided exploratory data analysis of the Gapminder data.
Using heuristics that account for fields' measure type and the dataset's keys, Umwelt generated smart default specifications to help the analyst quickly get started without needing to think deeply about low-level specification across three modalities.
Using an overview and detail strategy, Lula started by listening to the sonification, and contextualized what they heard with concrete data values by moving to the corresponding location in the textual hierarchy.
This process of smoothly moving between modalities allowed them to leverage the distinct affordances of each modality in a complementary way.
As they progressed, they recognized that their emergent goals during analysis would benefit from a change in the representations' affordances.
By making a small number of atomic changes in the editor, Lula was able to generate a new textual hierarchy and adjust the sonification playback order to explore the data a different way.

\begin{figure*}
  \centering
  \includegraphics[width=\textwidth]{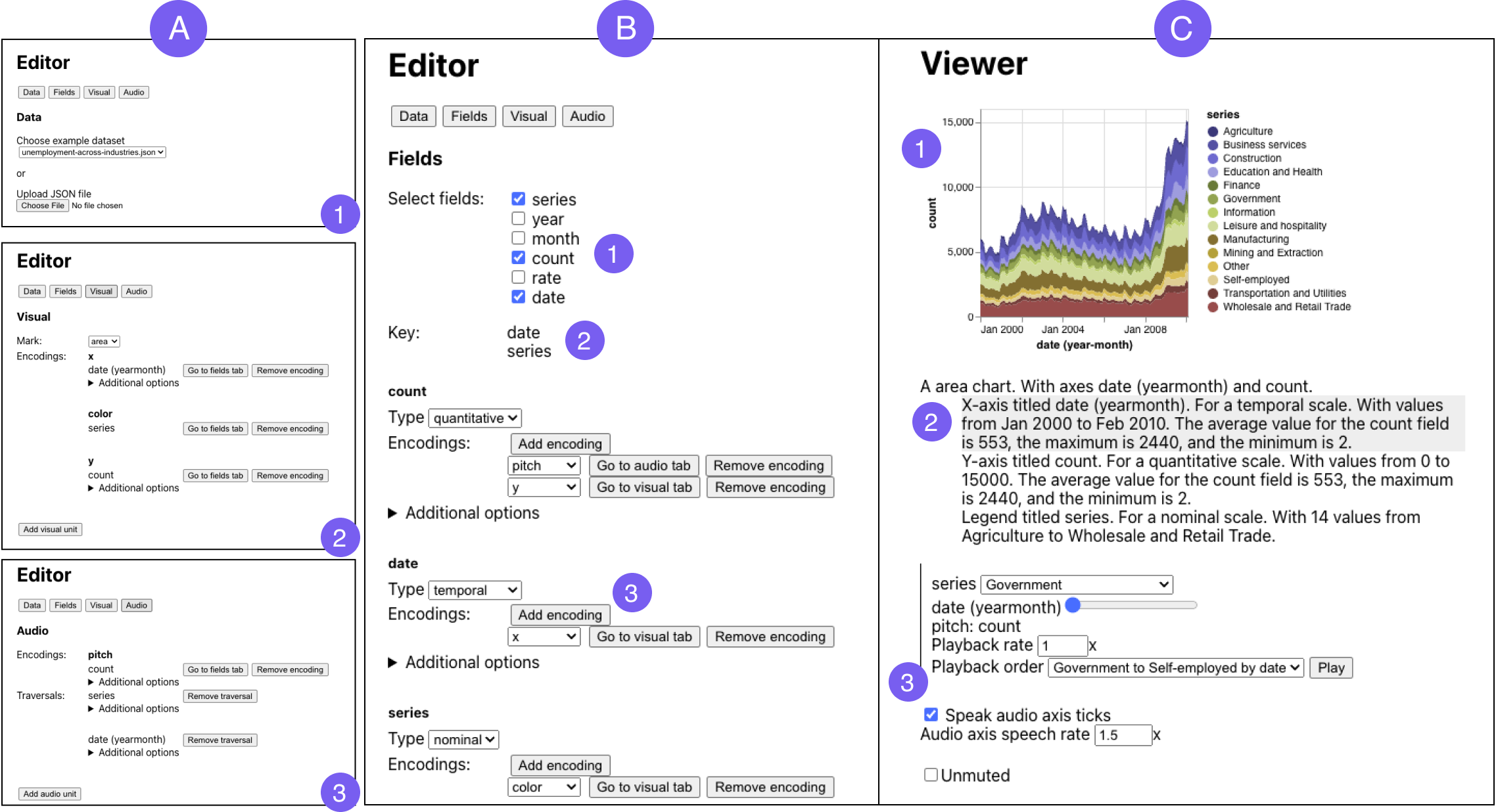}
  \caption{The Umwelt interface. A) The data, visual, and audio tabs of the editor. B) The editor's fields tab, where users specify field definitions and encodings. C) The viewer, where users analyze data with interactive multimodal data representations.}
  \Description{ A graphic with three sections that are labeled A through C. Each section includes labels 1-3. 

  Section A shows three Umwelt editor views corresponding to labels 1-3 in the following order: data, visual, and audio tabs. 

  Section B shows the editor on the fields tab. Label 1 is next to the checkboxes for the select fields option. Label 2 is next to the key description. Label 3 is next data encoding settings.

  Section C shows the viewer. Label 1 is next to a data visualization that reflects the settings in the editor's fields tab. Label 2 is next to the visualization's corresponding olli structure. Label 3 is next to the viewer's audio controls. 
  
  }
  \label{fig:interface}
\end{figure*}

\section{The Umwelt Editor}
\label{sec:editor}

\begin{figure*}
  \centering
  \includegraphics[width=\textwidth]{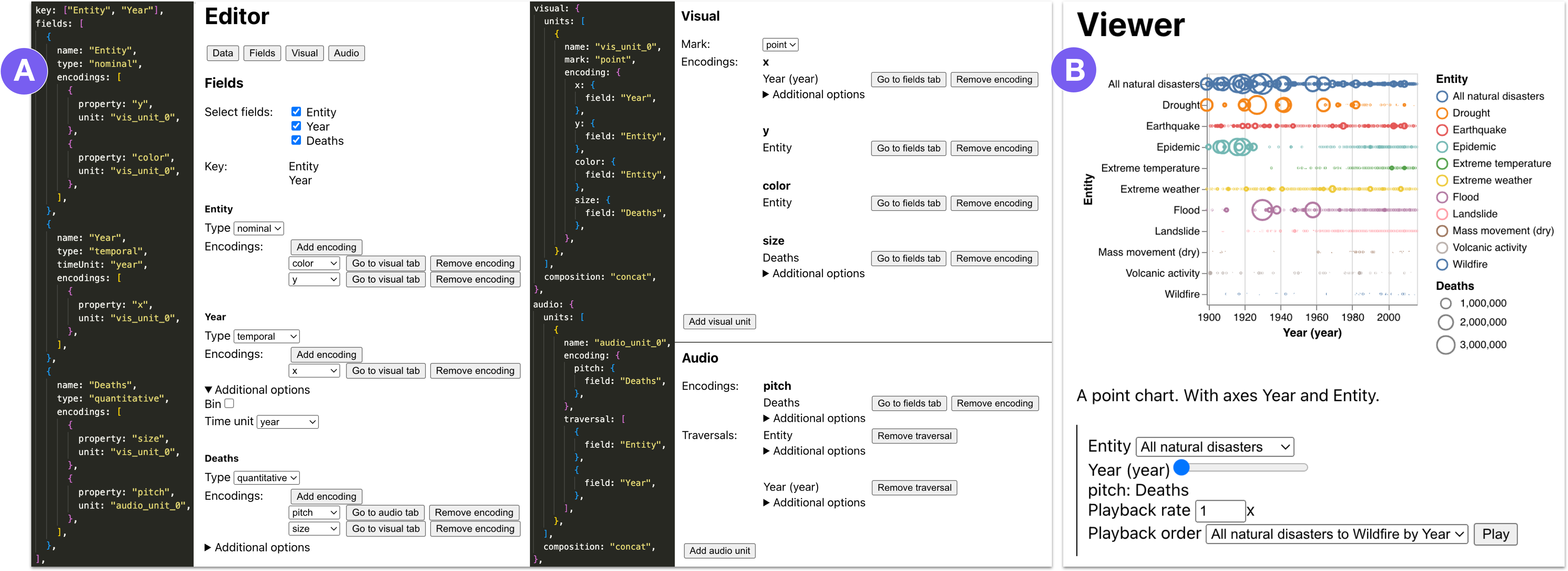}
  \caption{A) Fragments of an internal declarative specification shown next to their corresponding Umwelt editor states. B) The output multimodal representation for that specification.}

  \label{fig:spec-plus-editor}
  
  \Description{On the left, a declarative specification describes a dataset with three fields. Each field has a type definition and a list of encodings. There are two modality-specific specifications, visual and audio, which have the full definition of each encoding. On the right, a bubble plot with a textual description and a set of sonification controls generated by the specification.}
\end{figure*}

With Umwelt's editor, users specify multimodal representations using an interface designed primarily for screen readers.
The editor's internal state consists of a declarative JSON structure as shown in \autoref{fig:spec-plus-editor}.
In this section, we first introduce key parts of the editor, including its main components and its default specification heuristics. 
Then, we discuss our design rationale and how it addresses our design goals.

\subsection{Components of the Umwelt Editor}

Umwelt's editor, as shown in \autoref{fig:interface}A and \ref{fig:interface}B, is split across four tabs.
This organization is motivated by screen reader affordances, and navigation and wayfinding principles. In our co-design process, we found that when a screen reader user wanted to move back and forth between the editor and \viewer{} with their screen reader, it was more difficult to maintain their position if the editor had too much extraneous content visible at once. 
Organizing the interface into tabs helps screen reader users think about what functionality they need at a given moment, and helps manage page length and the depth of the information hierarchy.

\textbf{\textsc{Data} Tab (\autoref{fig:interface}.A.1).} 
A user begins by either loading a tabular dataset or choosing from a pre-populated list of example datasets. 
Umwelt then performs some simple type inference, and populates the other tabs with the dataset's fields.

\textbf{\textsc{Fields} Tab (\autoref{fig:interface}.B).} This tab lists all the fields in the dataset, with corresponding checkboxes to allow a user to pick which fields should participate in the multimodal representation.
When a user checks or unchecks a field, the system evaluates a set of heuristics (described in \ref{sec:default-specification}) to produce a default multimodal representation.
For each selected field, the editor provides a set of controls (\autoref{fig:interface}.B.3) to edit the field's inferred measure type, groupings, and transforms that may be calculated on the field (e.g. aggregation, binning).
These definitions serve as a shared default across all modalities\,---\,defaults that can be overridden in modality-specific ways under the appropriate tab (described below).
This tab also collates together the encodings a field is participating for both visual and audio modalities, offering user's a cross-modality perspective that can be important for generating cohesive and complementary experiences as we describe in \S~\ref{sec:field-encoding-spec}.

\textbf{\textsc{Visual} Tab (\autoref{fig:interface}.A.2).} This tab allows a user to make edits that apply only to the visual modality.
A visual specification includes the visual-specific concept of a mark, and the encodings for that modality that were assigned in the \textsc{Fields} tab. 
Changes to a field definition (e.g. its transforms) apply only to the corresponding visual encoding.
To allow users to be able to express multi-view displays (e.g., layered views or small multiples), Umwelt groups a mark and set of visual encodings into a \textit{visual unit}\,---\,a concept Umwelt inherits from Vega-Lite~\cite{satyanarayan_vega-lite_2017}.
Users can create multiple visual units, which can then be composed together as layers (where units are plotted one on top of the other) or as a concatenation (where units are laid out side-by-side).

\textbf{\textsc{Audio} Tab (\autoref{fig:interface}.A.3).} This tab allows a user to make edits that apply only to the audio modality.
An audio specification includes encodings from the \textsc{Fields} tab, which can be overridden, and \textit{traversals}, an audio-specific abstraction we introduce to control the order in which data points are sonified.

While some visualization systems such as Vega-Lite  offer an \textit{order} visual encoding channel (and Tableau offers similar functionality via its \textit{detail} shelf), this channel need only be used in special circumstances\,---\,for instance, to control the order that line segments are drawn as part of a connected scatterplot, or to determine z-axis and stack ordering. 
In contrast, ordering is much more central to the audio modality as data must be linearized into a fixed playback order, and
different orderings afford different data lookups and comparisons.
For instance, a reader of the stacked area chart in \autoref{fig:interface}.C may want to compare all values for \texttt{year}, one \texttt{series} at a time, to understand the trend of \texttt{count} within each \texttt{series}; or, they may want to traverse all \texttt{series} for a single \texttt{date} before moving onto the next \texttt{date}, to compare which series had the largest \texttt{count} at each \texttt{date}.
A visualization reader could easily do both of these operations on the same chart.
However, unlike the visual modality, a sonification can only afford one of these operations at a time --- switching between the two requires re-ordering the data.
Therefore an explicit specification of traversal is required.

A traversal specification is an ordered list of field definitions.
The ordered list represents the precedence of groupby operations over the data, which are used to determine a linearized playback order.
Consider the \autoref{fig:interface} example again.
The editor state in \autoref{fig:interface}A.3 defines a traversal \texttt{[series, date]}. 
This means that the data is first grouped by \texttt{series} before \texttt{date}.
In the corresponding viewer state in \autoref{fig:interface}C.3, when a user presses play, the sonification will select the first value of \texttt{series} and iterate through all values for \texttt{date}.
At each step, the current tuple of \texttt{(series, date)} values is used to query the value of \texttt{count} to encode it as pitch.
After traversing all \texttt{date} values for the given \texttt{series}, the playback advances to the next value for \texttt{series} and iterates over all \texttt{date} values again.
Consider the alternate traversal definition of \texttt{[date, series]}, which reverses the order in which the fields are grouped.
In this case, the sonification would instead start with the first value for \texttt{date} and iterate through all values for \texttt{series} before proceeding to the next \texttt{date}.
These two possible traversal specifications correspond to the two use cases described in the previous paragraph.

As this example demonstrates, it sometimes takes multiple sonification specifications to reproduce functionality afforded by a single visualization.
To make it easier to provide multiple alternate sonifications, we also extend the concept of view composition to sonification.
Like a unit visualization, a \textit{unit sonification} contains a single set of encodings and traversals.
Each audio unit corresponds to a single audio track that maps data to a tone, varying its properties (e.g. pitch, volume) according to the specified encodings.
Concatenating two unit sonifications means providing two separate, independent audio playback controls side-by-side.
A user can move between them to control which one they are listening to, and only one audio unit can be playing at once.
Layering two unit sonifications means that they share a traversal and that their encodings are expressed through two audio tones playing simultaneously.

\textbf{Future Work: Extending to Text and Other Modalities.}
Each modality is specified independently, yet each specification inherits from Umwelt's shared field definition.
As such, we expect it will be relatively straightforward to extend Umwelt's editor to support additional modalities (e.g., textual descriptions, tactile graphics, haptic feedback, etc.) provided there are well-defined abstractions and specification languages for these modalities.
For now, although Umwelt currently supports textual output, we have chosen to not offer a \textsc{Text} tab as this remains a nascent research area without settled consensus on suitable abstractions.
Instead, we have opted to 
preserve consistency between the textual structure and the visualization. We explain our rationale for this choice, and how it indicates a need for future research, in \ref{sec:aligned}.

\subsection{Default Specifications and Heuristics}
\label{sec:default-specification}

Umwelt uses a set of simple heuristics to generate default multimodal representations based on a dataset's typings and key.
In doing so, Umwelt seeks to avoid presenting a user with a blank slate whenever possible, and to accelerate a user in producing commonly used multimodal representations. 
Once the heuristics are evaluated, a user can modify the resultant defaults via the editor interface. 
These non-exhaustive heuristics are simple if-else statements that map combinations of field types and primary keys to specification fragments, which we document in Table \ref{tab:heuristics}.
In our example gallery (\autoref{fig:gallery}), examples A, B, C, G, and F were generated by Umwelt's heuristics while the rest required manual specification.

Our heuristics are motivated by the idea of \textit{functional dependence} between fields in a dataset.
In database theory, a functional dependence is the relationship described by a dataset's primary key --- a set of fields whose values uniquely index all rows of the dataset  \cite{codd_further_1971, munzner_visualization_2014}.
Just as search algorithms over relational databases use the key to perform data lookups \cite{codd_further_1971}, an analyst using a data representation will often implicitly use the key to look up functionally dependent fields (also known as value fields).
For example, a common way to read a single-series line chart is to choose a value for the x-axis field to look up a value for the y-axis field.
Though some existing visualization systems, including Tableau, use key-value relationships to model visualizations~\cite{stolte_polaris_2002}, this concept is even more central to Umwelt because it provides a shared basis for expressing a dataset's semantics across multiple modalities.

We identified and validated our heuristics through our co-design process, manually authoring specifications for a diverse range of datasets with differently arrangements of typings and key. 
We worked to identify commonalities between our designs and try to articulate our intuition for why certain representations made more sense than others.
For instance, we felt that the stacked area chart in \autoref{fig:interface}.C would be nonsensical if the \texttt{count} were encoded as \texttt{color}, and \texttt{series} as \texttt{y}. This can be explained by the functional dependence of \texttt{count} on \texttt{date} and \texttt{series}.
In the sonification case, the key constrains which fields should be encoded at all; we were almost always interested in mapping a value field to an encoding property like pitch and using the key to determine the order of playback.
For example, it does not make sense to sonify \texttt{date} or \texttt{series} in \autoref{fig:interface}.C\,---\,again, because \texttt{count} is functionally dependent \texttt{date} and \texttt{series}.
Finally, in the case of text, the key imposes constraints on the hierarchical structure.
In \autoref{fig:interface}.C, we were more likely to want to group by \texttt{date} and \texttt{series} to look up a \texttt{count} value than, for example, to look up a \texttt{date} by first finding its corresponding \texttt{count}.

\clearpage
\onecolumn
\begin{small}
\begin{longtable}{@{\extracolsep{\fill}} p{0.1\textwidth} p{0.05\textwidth} p{0.2\textwidth} p{0.2\textwidth} p{0.25\textwidth} p{0.08\textwidth}}
\caption{ Default specification heuristics based on a dataset's key and typings. T = temporal field, N = nominal field, Q = quantitative field. Each row represents a rule that matches a dataset's key and value tuples. The visualization, sonification, and textual description columns show default specifications for each rule, represented in YAML format for conciseness.} \label{tab:heuristics} \\
  \toprule
  Key & Value & Visualization & Sonification & Textual Structure & Example \\
  \midrule
  T, N & Q & 
  \vspace{-16pt}
  \begin{lstlisting}[language=yaml]
mark: "line"
encoding:
  x:
    field: t_key
  y:
    field: value[0]
  color:
    field: n_key
  \end{lstlisting} 
  \vspace{-16pt}
  & 
  \vspace{-16pt}
  \begin{lstlisting}[language=yaml]
encoding:
  pitch:
    field: value[0]
traversal:
  - field: n_key
  - field: t_key
  \end{lstlisting} 
  \vspace{-16pt}
  & 
  \vspace{-16pt}
  \begin{lstlisting}[language=yaml]
groupby: n_key
children:
  - groupby: t_key
  - groupby: value[0]
  \end{lstlisting} 
  \vspace{-16pt}
  & \autoref{fig:gallery}A \\
  \midrule
  T, N (>5 categories)& Q & 
  \vspace{-16pt}
  \begin{lstlisting}[language=yaml]
mark: "point"
encoding:
  x:
    field: t_key
  y:
    field: n_key
  color:
    field: n_key
  size:
    field: value[0]
  \end{lstlisting} 
  \vspace{-16pt}
  & 
  \vspace{-16pt}
  \begin{lstlisting}[language=yaml]
encoding:
  pitch:
    field: value[0]
traversal:
  - field: n_key
  - field: t_key
  \end{lstlisting} 
  \vspace{-16pt}
  & 
  \vspace{-16pt}
  \begin{lstlisting}[language=yaml]
- groupby: t_key
- groupby: n_key
- groupby: value[0]
  \end{lstlisting} 
  \vspace{-16pt}
  & \autoref{fig:gallery}G \\
  \midrule
  -- & Q, Q, N & 
  \vspace{-16pt}
  \begin{lstlisting}[language=yaml]
mark: "point"
encoding:
  x:
    field: q_value[0]
  y:
    field: q_value[1]
  color:
    field: n_value
  \end{lstlisting}
  \vspace{-16pt}
  & 
  \vspace{-16pt}
  \begin{lstlisting}[language=yaml]
- encoding:
    pitch:
      field: value[0]
      aggregate: "mean"
  traversal:
    - field: value[1]
      bin: true
- encoding:
    pitch:
      field: value[1]
      aggregate: "mean"
  traversal:
    - field: value[0]
      bin: true
  \end{lstlisting} 
  \vspace{-16pt}
  & 
  \vspace{-16pt}
  \begin{lstlisting}[language=yaml]
- groupby: q_value[0]
- groupby: q_value[1]
- groupby: n_value
  \end{lstlisting} 
  \vspace{-16pt}
  & \autoref{fig:gallery}B \\
  \midrule
  T & Q, Q & 
  \vspace{-16pt}
  \begin{lstlisting}[language=yaml]
mark: "line"
encoding:
  x:
    field: value[0]
  y:
    field: value[1]
  order:
    field: key[0]
  \end{lstlisting} 
  \vspace{-16pt}
  & 
  \vspace{-16pt}
  \begin{lstlisting}[language=yaml]
- encoding:
    pitch:
      field: value[0]
  traversal:
    - field: key[0]
- encoding:
    pitch:
      field: value[1]
  traversal:
    - field: key[0]
  \end{lstlisting} 
  \vspace{-16pt}
  & 
  \vspace{-16pt}
  \begin{lstlisting}[language=yaml]
- groupby: key[0]
- groupby: value[0]
- groupby: value[1]
  \end{lstlisting} 
  \vspace{-16pt}
  & \autoref{fig:gallery}C \\
  \midrule
  T, N, N & Q & 
  \vspace{-16pt}
  \begin{lstlisting}[language=yaml]
mark: "point"
encoding:
  x:
    field: value[0]
  y:
    field: n_key[1]
  color:
    field: t_key
  facet:
    field: n_key[0]
  \end{lstlisting} 
  \vspace{-16pt}
  & 
  \vspace{-16pt}
  \begin{lstlisting}[language=yaml]
encoding:
  pitch:
    field: value[0]
traversal:
  - field: n_key[0]
  - field: n_key[1]
  - field: t_key
  \end{lstlisting} 
  \vspace{-16pt}
  & 
  \vspace{-16pt}
  \begin{lstlisting}[language=yaml]
groupby: n_key[0]
children:
  - groupby: value[0]
  - groupby: n_key[1]
  - groupby: t_key
  \end{lstlisting} 
  \vspace{-16pt}
  & \autoref{fig:gallery}F \\
  \midrule
  T, N & Q, Q & 
  \vspace{-16pt}
  \begin{lstlisting}[language=yaml]
mark: "line"
encoding:
  x:
    field: value[0]
  y:
    field: value[0]
  facet:
    field: n_key
  color:
    field: n_key
  order:
    field: t_key
  \end{lstlisting} 
  \vspace{-16pt}
  & 
  \vspace{-16pt}
  \begin{lstlisting}[language=yaml]
- encoding:
    pitch:
      field: value[0]
  traversal:
    - field: n_key
    - field: t_key
- encoding:
    pitch:
      field: value[1]
  traversal:
    - field: n_key
    - field: t_key
  \end{lstlisting} 
  \vspace{-16pt}
  & 
  \vspace{-16pt}
  \begin{lstlisting}[language=yaml]
groupby: n_key
children:
  - groupby: value[0]
  - groupby: value[1]
  - groupby: t_key
  \end{lstlisting} 
  \vspace{-16pt}
  & \autoref{fig:walkthrough}.7 \\
  \bottomrule
\end{longtable}
\end{small}
\twocolumn
\clearpage
\vspace{-12pt}

\begin{figure*}[t!]
  \centering
  \includegraphics[width=\textwidth]{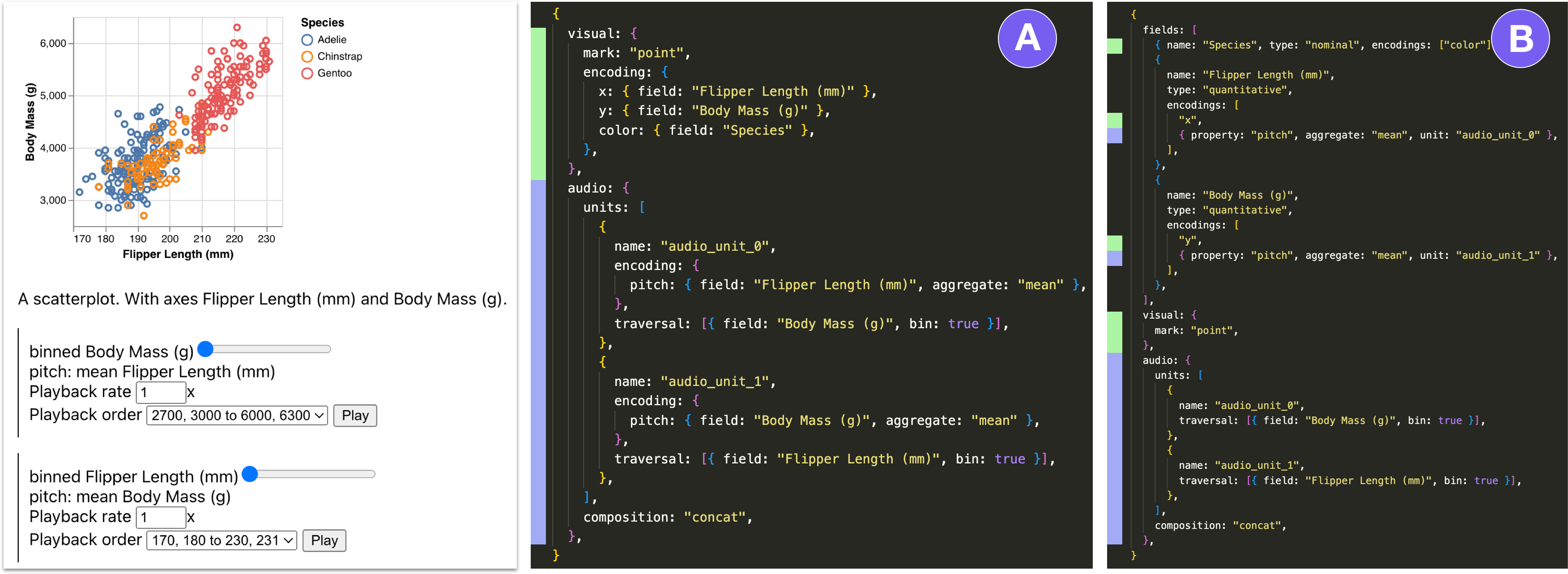}
  \caption{Prototypes of encoding- and field-oriented specifications of a scatterplot with concatenated sonification, illustrating the \textit{role-inexpressiveness} \cite{blackwell_cognitive_2001} of field-oriented textual specification. Color-coded spans on the left side of each text prototype show the lines of code that pertain to each modality: green represents visual while blue represents audio. A) Encoding-oriented specification groups each modality into unit specifications. B) Field-oriented specification fragments each unit's encodings across the spec.}

  \label{fig:orientation}
  
  \Description{A figure divided into three parts. The first part is a screenshot of the Umwelt viewer, representing a penguin dataset using a scatterplot, a textual description, and two sets of sonification controls. The scatterplot has Flipper Length on the x-axis, Body Mass on the y-axis, and Species on the color legend. The first set of sonification controls maps mean Flipper Length to pitch, and traverses it with binned Body Mass. The second set is the other way around, mapping mean Body Mass to pitch and traversing with binned Flipper Length.
  
  The second and third parts of the figure are labeled A and B and contain JSON code snippets showing how to express the aforementioned multimodal data representations via encoding-oriented and field-oriented specifications, respectively.}
\end{figure*}

\subsection{Design Rationale}

\subsubsection{Field- vs Encoding-Oriented Specification}
\label{sec:field-encoding-spec}

The design of a specification language can impose constraints on a user's order of operations \cite{blackwell_cognitive_2001}.
Conventional visualization grammars, including Vega-Lite, are \textit{encoding-oriented}: encoding is a top-level abstraction in a Vega-Lite unit specification, and field definitions are nested within encodings.
However, a consequence of encoding-oriented specification is that users must first decide what encodings they are using before assigning fields to them, requiring them to have visual idioms in mind when initially formulating their goals.
This limitation is even more pronounced in the context of multimodal representations, as an author may not even have an initial choice of modality in mind.
When we began designing Umwelt, we first designed it as an encoding-oriented declarative JSON language.
However, our co-design process led us to explore \textit{field-oriented} specification as an alternative: fields are top-level entities and encoding definitions are nested within fields.
We felt that a field-oriented approach was amenable to multimodal authoring because an author can make localized changes to a single field definition used across multiple modalities, or switch a field's encoding from one modality to another.
The increased ease of these changes enables deferred commitment to a specific representation (DG1).
Comparing encoding-oriented and field-oriented specification using the cognitive dimensions framework \cite{blackwell_cognitive_2001}, we argue that field-oriented specification reduces \textit{viscosity} (difficulty of making changes) and increases \textit{provisionality} (ease of exploratory prototyping).

Although field-oriented specification helped us address one of our design goals, we found that it became much more difficult to understand a specification without using modality-specific abstractions.
Encoding-oriented specification is prevalent amongst existing declarative grammars because its syntax captures an important semantic property of the relationship between encodings and fields --- namely, that each encoding property can only have one field assigned to it.
This in turn enables the concise expression of other top-level abstractions: for instance, a unit visualization has one mark and one set of encodings.
When reading a Vega-Lite spec, it is easy to understand that a mark and a set of encodings are associated together because they are contained within the same unit spec.
When we switched to a field-oriented language, we found that modality-specific definitions became fragmented across field definitions.
Consider the example in \autoref{fig:orientation}.
The encoding-oriented specification in \ref{fig:orientation}A uses unit specs to group the functionality of each modality together.
But in \ref{fig:orientation}B's field-oriented specification, encodings belonging to the same unit specification are nested under multiple field definitions.
Further, in \ref{fig:orientation}B, modality-specific concepts like mark or traversal are not nested under any individual field, so additional verbosity or repetition must be introduced to associate these concepts with their respective units.
In terms of cognitive dimensions, field-oriented specification introduces \textit{role-inexpressiveness} \cite{blackwell_cognitive_2001} because it is more difficult to read a specification and clearly understand relationships and dependencies between entities affecting the same modality.

Field-oriented and encoding-oriented approaches both had affordances that felt essential but were in conflict with each other in a textual language, leading to significant tension in our design process.
Our co-design process led us to bridge between field- and encoding-oriented specification by designing Umwelt primarily as a structured editor interface, rather than as a textual JSON language.
In the editor, the \textsc{Fields} tab allows a user to create a field-oriented specification by populating a field with encodings from any modality.
Then, the user can switch to the \textsc{Visual} or \textsc{Audio} tabs to edit modality-specific attributes like mark and traversal, or perform actions that are scoped to one modality (like adding or removing a unit spec).
In its internal representation (shown in \autoref{fig:spec-plus-editor}), Umwelt maintains both field-oriented and encoding-oriented abstractions.
It links the two kinds of specification together via references.
In our prototype language designs, expressing these references in a textual specification language was unwieldy and lead to repetition, but they are suitable for an interface where a user can easily navigate between two views into the same underlying spec.
Our eventual design for Umwelt prioritizes field-oriented specification to encourage ease of switching between modalities during exploratory specification, but also allows users to switch to encoding-oriented specification for detailed control.

Designing Umwelt as a structured editor also introduces additional benefits.
An editor interface can reactively update the options it presents to a user based on its current state and can, thus, hide operations that would lead a user to invalid states.
As a result, each atomic edit in the editor is a transition from one valid specification to another.
In contrast, with a textual language, any time a user is partway through typing out a statement, the program will not compile.
In terms of the cognitive dimensions of notation framework \cite{blackwell_cognitive_2001}, we would say that an editor interface reduces \textit{error-proneness} compared to the textual language, and affords users a better ability to \textit{progressively evaluate} \cite{blackwell_cognitive_2001} the specification they are editing.

\subsubsection{Aligned vs Disjoint Modalities}
\label{sec:aligned}

In a multimodal data representation, how each representation relates to the others can reflect different design priorities.
For example, modalities can be \textit{aligned} in that they redundantly encode the same information, emphasizing a cohesive insight or set of possible comparisons.
Or, modalities can be \textit{disjoint}, conveying different aspects of the data that can be synthesized together into a greater whole than the message of each individual representation.
In existing systems that derive non-visual representations from the visual, the derived representations are inherently aligned with the original.
But in systems like Umwelt where modalities are independent, it can be up to the author's discretion whether modalities are aligned or disjoint.

In our co-design process, thinking about aligned and disjoint modalities uncovered a tension in our design goals, where we seemingly could not simultaneously prioritize DG2 and DG3.
On one hand, using visualization and sonification as \textit{disjoint} modalities meant that we could use sonification to focus on comparisons between fields that are difficult to compare visually, or encode fields that are not present in the chart because it would be too visually overwhelming to include them.
This additional expressiveness contributes to DG2, where a user can gain additional information from the use of multiple modalities together.
On the other hand, using visualization and text as \textit{aligned} modalities preserves consistency between the two representations, which is crucial for BLV users who need the textual representation to access the visualization.
During the authoring process, a screen reader user needs the representations to align to verify that they are creating sensible visuals. This verification process is crucial to DG3, establishing common ground between blind and sighted users.

In sum, the textual structure can serve a dual purpose of (1) textually conveying the data and (2) making the visualization accessible.
These two purposes fulfill DG2 and DG3, respectively, but it is difficult to fulfill both purposes simultaneously because one implies a disjoint representation while the other implies an aligned representation.
While Umwelt could allow authors to override field definitions in the textual modality, this would cause the visualization and textual description to become disjoint.
Our co-design process led us to prioritize aligned visual and textual representations, and we made a decision not to expose a \textsc{Text} tab in the editor.
Nonetheless, disjoint visual and textual representations is an important area for future design exploration.
For instance, researchers could explore ways to enable a user to customize whether a textual structure is aligned or disjoint on-the-fly.

\section{The Umwelt Viewer}
\label{sec:viewer}

Umwelt's \viewer{} renders interactive multi-modal representations specified in the editor, including a visualization, a structured textual description, and a sonification.
In this section, we first introduce the \viewer's components and its linked interaction model. 
Then, we discuss our design rationale and how the \viewer{} addresses our design goals.

\subsection{Multi-Modal Data Representations}

The Umwelt Viewer, as shown in \autoref{fig:interface}C, consists of three components: a visualization, a textual structure, and a sonification. Though there is no explicit interaction specification in the editor, each representation in the \viewer{} is implicitly interactive. 
This interaction-first approach to the design of the \viewer{} is motivated by the need to selectively attend to data.
Interactive representations enable a user to select a subset of data and share that selection across multiple representations.
Here, we describe each representation before discussing their interactive behavior in more detail in \ref{sec:linked-interaction}.

\textbf{Visualization.}
Umwelt converts its internal representation into a Vega-Lite \cite{satyanarayan_vega-lite_2017} specification to render a visualization (\autoref{fig:interface}.C.1).
It augments this specification with additional Vega-Lite selection parameters, resulting in a visual representation that is interactive by default. 
For example, a user can drag on the visualization to select a rectangular region of data.

\textbf{Textual structure.}
Umwelt renders a structured textual description (\autoref{fig:interface}.C.2) with Olli \cite{blanco_olli_2022}, an open-source library that implements Zong, Lee, Lundgard et al.'s design dimensions for screen reader experiences \cite{zong_rich_2022}.
Olli outputs a hierarchical structure in the shape of a tree.
Each node in the structure is associated with a textual description.
The root of the structure gives an overall description of the data, while deeper levels in the structure apply successive filters on the data to give more granular descriptions.

The textual output does not require a visual specification, but can use visual information to augment its structure and descriptions.
When there is a visual specification present, Olli structures the tree based on the visualization's encodings, and reference visual concepts in its description.
As we discussed in \ref{sec:aligned}, this makes the visual representation accessible for screen reader users and establishes common ground.
On the other hand, when there is no visual specification, Olli outputs a relatively flat structure that allows a user to group the data by each field, and uses descriptions that do not reference visual concepts.

\textbf{Sonification.}
Umwelt implements an interactive sonification runtime to render its audio specifications (\autoref{fig:interface}.C.3).
For each audio unit specification, Umwelt renders a set of audio controls representing a single audio track.
A user can press the play button (or the `p' key on their keyboard) to play and pause the sonification.
They can also interactively control their position within the sonification using input elements (i.e., dropdown menus for nominal and ordinal fields, and sliders for quantitative and temporal fields).

To help users keep track of their position in the sonification playback, the sonification runtime uses spoken announcements of data values interleaved with the sonification to communicate playback progress (exposed in the interface as an option called ``audio axis ticks'').
For example, in \autoref{fig:interface}C, as the sonification traverses \texttt{date} values, the system will speak the \texttt{date} value (e.g. Jan 2000) before playing the sonified segment between each axis tick.
If a visualization is present, these ticks will always correspond to the visual ticks for consistency.
Otherwise, they are determined by binning the key fields' domains to calculate regularly spaced intervals (or, for categorical fields, directly reading the value corresponding to each category).

Playback order is an important consideration for sonification, because different playback orders can facilitate different comparisons. For instance, in \autoref{fig:interface}C, playing through all \texttt{date} values for a given \texttt{series} before advancing to the next \texttt{series} is akin to the visual operation of reading each line left to right, getting a sense of each line's trend.
On the other hand, playing through all \texttt{series} for a given \texttt{date} before advancing to the next \texttt{date} is akin to vertically comparing the values for a given x-axis position.
Because the preferred order will depend on a user's goals, Umwelt determines the initial playback order by the ordering of the traversal specification and provides a dropdown menu to select an alternate playback order.

\subsection{Coordinating Modalities with Linked Interactions}
\label{sec:linked-interaction}

\begin{figure*}
  \centering
  \includegraphics[width=\textwidth]{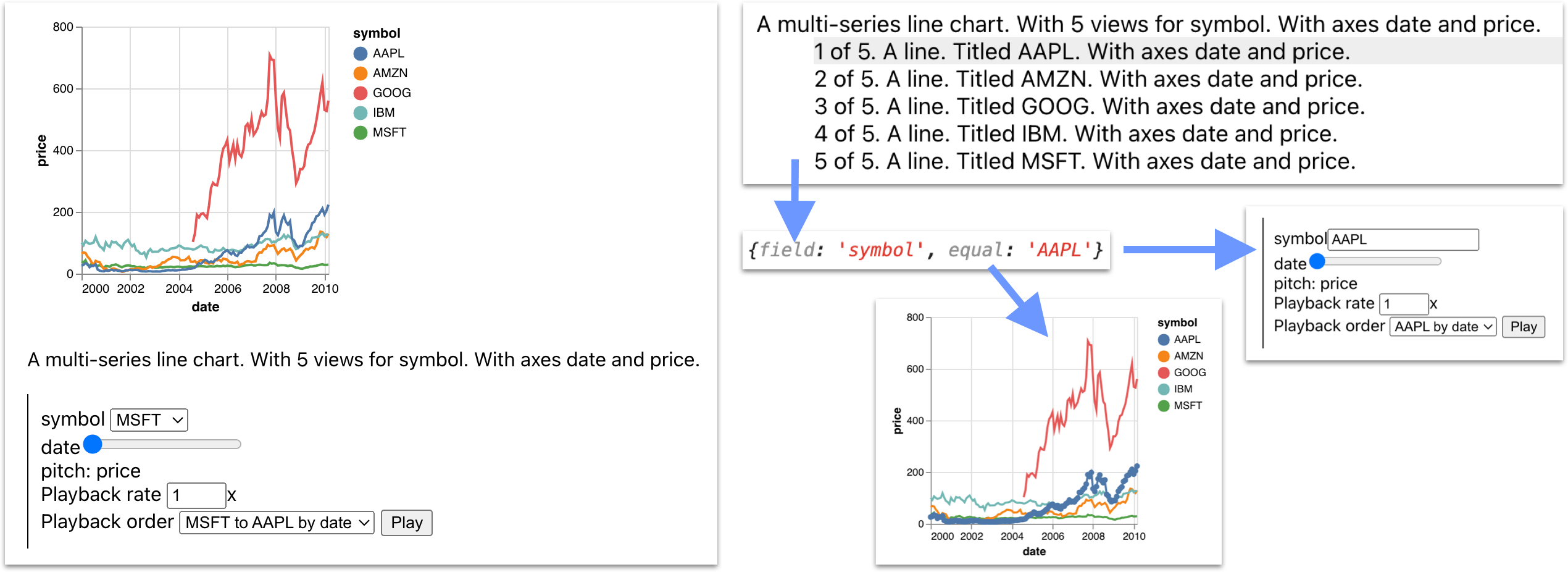}
  \caption{An example of linked interaction across modalities, driven by the textual modality. Navigating to a node in the textual structure emits a query predicate. The visualization reifies this predicate as a conditional encoding, and the sonification reifies it as a filter.}

  \label{fig:predicate}
  
  \Description{The left side of the figure shows a multi-series line chart, a textual description, and a set of sonification controls. The right side shows the process of linked interactions step by step. First, a user navigates to a location in the textual structure describing a line titled AAPL. This emits the following predicate: \{field: 'symbol', equal: 'AAPL'\}. This predicate is sent to the visualization, which highlights the corresponding line. It is also sent to the sonification, which filters its domain down to only include the data matching AAPL.}
\end{figure*}

Though each modality maintains its own interactive state, Umwelt links interactions across modalities to aid analysis.
Each modality has one or more interactions that define a selection over the data, and can be modeled as query predicates.
For instance, a user can drag a brush over the visualization, navigate to a location or define a custom filter in the textual structure, and navigate to a position in the sonification playback.
When a user performs one of these interactions on a representation, that representation updates its own state and emits a query predicate to the other representations.
Each representation then reifies this predicate as some sort of effect (e.g. a transformation).
\autoref{fig:predicate} shows an example of this process, driven by the textual modality.
Olli associates a query predicate with each node in its structure --- as a user navigates through the structure, the current node's predicate describes the data selected by the user's current position.
In this example, a user navigates to a node corresponding to the predicate \texttt{\{field: \textquotesingle{symbol}\textquotesingle, equal: \textquotesingle{AAPL}\textquotesingle\}}.
This interaction emits the predicate to the visualization and the sonification.
The visualization updates to visually highlight the selected data, and the sonification filters its domain to match the selected data.

\subsection{Design Rationale}

\subsubsection{Highlighting vs Zooming in Non-Visual Representations}

In visualization, the same user interaction could plausibly map to multiple possible effects.
For example, dragging a rectangular area on a Vega-Lite scatterplot could \textit{highlight} the data by giving it a conditional encoding (e.g., showing highlighted points in a different color).
Or, that same drag interaction could \textit{zoom} into that data (e.g. in an overview + detail interaction), resizing the viewport to only contain the selected data.

Analogously in non-visual modalities, there are multiple possible techniques for conveying the result of an interaction.
Consider an example in the textual modality, which we surfaced while prototyping ways to filter a textual structure.
One way of applying a filter to a textual structure is to re-scale the structure to fit the filtered data.
For instance, an x-axis that originally represented a domain of 0--100 by splitting it into five nodes representing increments of 20 might be re-scaled to split a filtered domain of 50--70 into four nodes representing increments of 5.
On testing this approach, co-designer \censor{Hajas} compared this feature to ``zooming in'' on a visualization by changing its viewport.
Another way of applying a filter is to leave the structure unchanged while re-flowing the structure with only the filtered data.
For instance, the previous example would still have five nodes representing increments of 20, but many of the nodes would be empty after applying the filter.
This approach is more analogous to ``highlighting'' a visualization, because the viewport remains the same but the un-selected data is de-emphasized.

Though zooming and highlighting appear to be recurring concepts across modalities, it is not clear that either is universally preferable.
Currently, Umwelt's visual representation uses highlighting to convey interactive state because this is a more common interaction technique in visualization.
This makes sense when considering the fact that visually, it is helpful to maintain a consistent viewport to situate a highlighted subset within the broader context of the full data.
However, in our co-design process, we felt that the ``zoom'' interaction made more sense as a default for text, since a structure that is not scaled to the data often requires a user to navigate through extraneous nodes to find useful data.
Guided by DG2, we chose these defaults per-modality according to each modality's affordances.
We also considered cases where the modalities are used together --- for instance, a sighted collaborator brushing on the visualization to momentarily draw a screen reader user collaborator's attention to a subset of data.
However, our choice of default potentially trades off consistency across modalities --- an important consideration for DG3.

Future work on interaction design for multi-modal data representations can work towards a better understanding of what types of approaches are best suited for certain situations or tasks, and how an author or end user might be able to switch between interaction techniques.
And, though we conducted this initial exploration in the textual modality, future work remains to explore how interaction concepts like conditional encoding and viewport scaling extend to other non-visual modalities, like interactive sonification, in the context of a multi-modal system.

\subsubsection{Preserving Interactive Context Across Modalities}

Because the representations are designed to be used together, we wanted to enable users to smoothly switch between modalities to facilitate complementary use (DG2).
This required us to think about how to maintain context when switching representations.
In our initial explorations, a co-author compared the ability to select data via navigation in the textual structure with ``pointing at part of a chart.''
We designed linked interactions so that the system could express a consistent understanding of what data the user is ``pointing'' at across all representations.

Another important goal of linked interaction was to establish common ground for collaboration and presentation (DG3).
One of the most important uses of data is to communicate with others, and not everyone in a conversation may use the same sensory modalities.
This is why, despite primarily designing Umwelt with screen readers in mind, we found it important to include a visual representation that visually conveys the state of a screen reader user's exploration in the textual structure or sonification.
Conversely, the textual structure and sonification update to reflect interactions on the visualization.
This also helps users think of the representations as complementary, e.g. by using one for wayfinding and the other for consuming \cite{jones_customization_2023} (DG2).

\section{Evaluation: Example Gallery}
\label{sec:examples}

\begin{figure*}
  \centering
  \includegraphics[width=\textwidth]{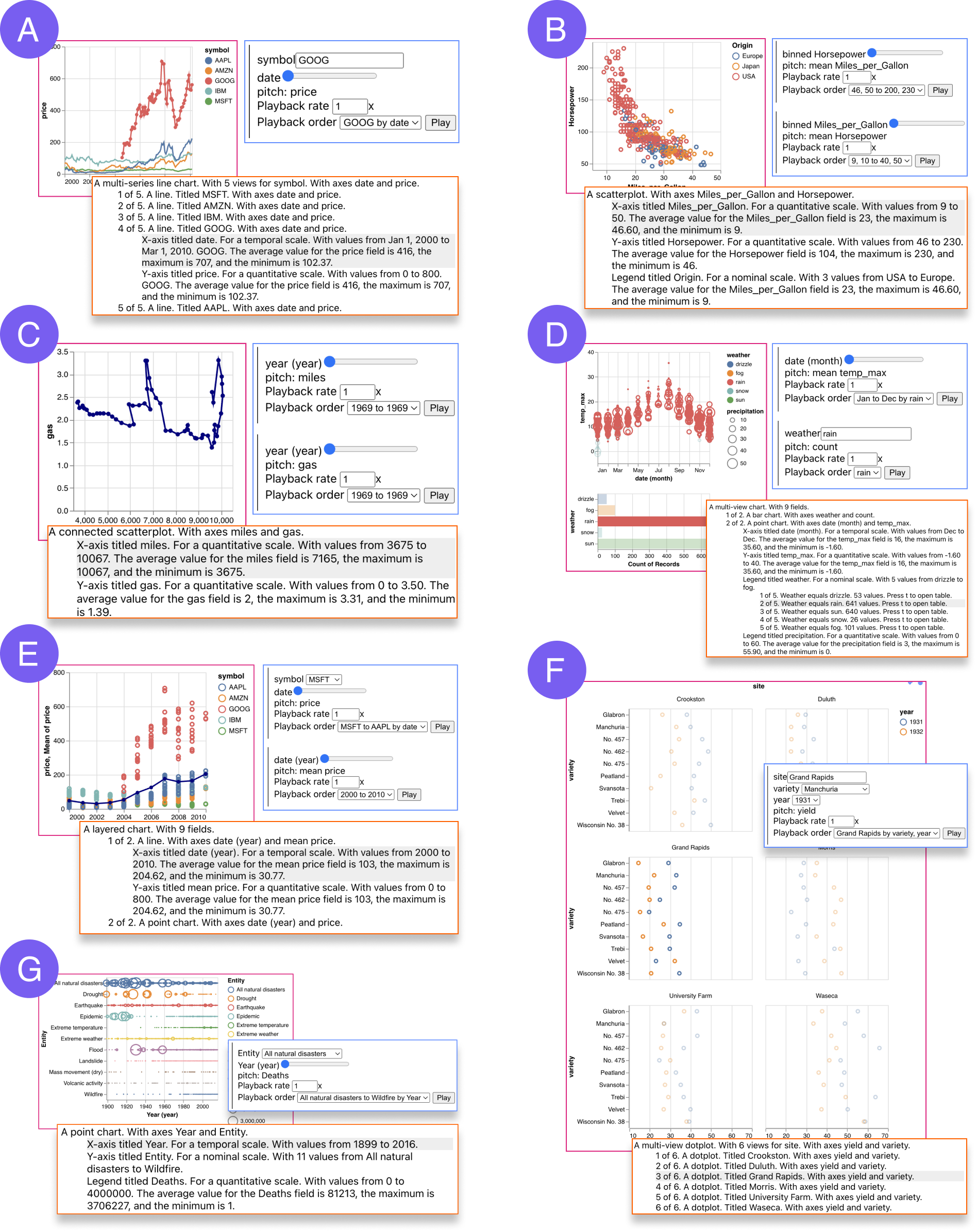}
  \caption{Example multi-modal representations created with Umwelt, expressing a range of key-value semantics. A) Two keys and one value. B) No keys and three values. Audio units represent the quantitative values' distribution in 2 dimensions. C) One key and two values. Audio units represent looking up each value with the same key. D) Concatenated visualization and sonification. E) Layered visualization with concatenated sonification. F) Three keys and one value. G) An alternate visual representation for the same key-value semantics as A.}

  \label{fig:gallery}
  
  \Description{ A graphic with seven sections that are labeled A through G. Each section includes a data visualization and it's corresponding Umwelt sonification controls and Olli structure.
  
  Section A's visualization is a multi-series line chart showing stock prices from 0 to 800 on the Y-axis against dates from 2000 to 2010 on the X-axis for AAPL, AMZN, GOOG, IBM, MSFT. The multi-series line chart has the line for GOOG bolded. The Umwelt sonification is focused on the symbol GOOG, with pitch being encoded by price and the playback order being GOOG by date. The Olli structure is focused on the X-axis for the line titled GOOG. 

  Section B's visualization is a scatterplot showing horsepower from 0 to 250 on the Y-axis against miles per gallon from 0 to 50 on the X-axis. The colors of the dots on the scatterplot represent the origin, with it either being of European, Japanese, and United States origin. The Umwelt sonification is focused on two items: the binned horsepower, with pitch being encoded by mean miles per gallon and the playback order being 46, 50 to 200, 230 and the binned miles per gallon, with pitch being encoded by mean horsepower and the playback order being 9, 10 to 40, 50. The Olli structure is focused on the X-axis titled miles per gallon.

  Section C's visualization is a connected scatterplot showing gas from 0 to 3.5 on the Y-axis against miles from 0 to 10,000 on the X-axis. The Umwelt sonification is focused on two items: year, with pitch being encoded by miles and the playback order being 1969 to 1969 and year, and year, with pitch being encoded by gas and the playback order being 1969 to 1969. The Olli structure is focused on the X-axis titled miles.

  Section D's visualization shows two plots. The plot on the top is a bubble plot showing max temperature from 0 to 40 on the Y-axis against months from January to December on the X-axis. The bubbles' colors represent different weathers such as drizzle, fog, rain, snow, and sun and the bubbles' sizes represent different precipitations such as 10, 20, 30, 40, and 50. The bubbles of color red, for rain, are bolded. The plot on the bottom is bar chart showing weather on the Y-axis for drizzle, fog, rain, snow, and sun against count of records from 0 to 700 on the X-axis. The bar for rain is bolded. The Umwelt sonification is focused on two items: month, with pitch being encoded by mean temp max and the playback order being Jan to Dec by rain and weather set to rain, with pitch being encoded by count and the playback order being rain. The Olli structure is focused on the legend where weather equals rain.

  Section E's visualization is a dotplot with layered average showing stock prices and mean of prices from 0 to 800 on the Y-axis against dates from 2000 to 2010 on the X-axis for AAPL, AMZN, GOOG, IBM, MSFT. The Umwelt sonification is focused on two items: symbol set to MSFT, with pitch being encoded by price and the playback order being MSFT to AAPL by date and date, with pitch being encoded by mean price and the playback order being 2000 to 2010. The Olli structure is focused on the X-axis titled year.

  Section F's visualization is a trellis plot, showing dotplots for different sites such as Crookston, Duluth, Grand Rapids, Morris, Univerity Farm, Waseca. Each dotplot shows different varieties such Glabron, Manchuria, No. 457, No. 462, No. 475, Peatland, Svansota, Trebi, Velvet, Wisconsin No. 38 on Y-axis for the years 1931 and 1932. The dotplot titled Grand Rapids is bolded. The Umwelt sonification is focused on the site, variety, and year, all of which are set to Grand Rapids, Manchuria, and 1932, respectively. Pitch is encoded by yield and the playback order is Grand Rapids by variety, year. The Olli structure is focused on the dotplot titled Grand Rapids.

  Section G's visualization is a bubble plot showing weather entities on the Y-axis against years from 1900 to 2020 on the X-axis. The bubbles' colors represent different weather entities such as All natural disasters, Drought, Earthquake, Epidemic, Extreme Temperature, Extreme Weather, Flood, Landslide, Mass movement (dry), Volcanic activity, Wildfire and the bubbles' sizes represent different amounts of deaths such as 1,000,000, 2,000,000, 3,000,000. The Umwelt sonification is focused on the entity All natural disasters, with pitch being encoded by deaths. The playback order is All Natural disasters to Wildfire by year. The Olli structure is focused on the X-axis titled year.
  
  }
\end{figure*}

To evaluate our approach's expressive extent, we used Umwelt to create a gallery of diverse multi-modal examples representing a variety of key-value semantics (\autoref{fig:gallery}). 
In addition to simple examples consisting of one visualization, one audio track, and one textual structure (A, G),
Umwelt provides a simple view composition abstraction that can express concatenated and layered visualizations (D, E), concatenated sonifications (B, C, D, E), and multi-view textual tree structures (A, D, E, F).

In contrast to prior approaches, such as Chart Reader \cite{thompson_chart_2023}, that were constrained to a small set of chart forms (and therefore key-value semantics), Umwelt can express more complex relationships among fields.
For instance, \autoref{fig:gallery}.B shows a dataset with an empty key and a set of values with two quantitative fields and one nominal field, visually represented as scatterplot.
Because there are two quantitative values, a user could plausibly want to look up either one by pitch.
\autoref{fig:gallery}.B provides two audio units so that users can choose which value field to sonify.
However, since there is no key by which to look either value up, \autoref{fig:gallery}.B's sonification uses binning and aggregation to transform the fields, creating a 1:1 correspondence between bins and aggregated values so that users can traverse the bins to look up a value.
The result is a 2 dimensional sonification that conveys the distribution of quantitative values in the x or y orientations.

Though this is not the only possible way to sonify a scatterplot, \autoref{fig:gallery}.B is illustrative of the importance of decoupling visual and non-visual specification (DG1) in order to express representations that achieve complementary goals (DG2).
A system that derives audio encodings from visual encodings might re-map the \texttt{x} and \texttt{y} encodings directly to \texttt{pitch}.
A user might want to bin before sonification, as shown in \autoref{fig:gallery}.B, in order to get a high level sonic overview of how the data is distributed along an axis without being overwhelmed by the fluctuating values of each individual data point.
But in a visualization-first system, because the visualization is not binned, a user would not be able to add binning to the sonification without first changing the visualization to a binned representation (e.g. heatmap).
Yet, the user may not want to align the modalities in this way; they may want to switch between the sonification overview and a visual or textual representation of individual data points.
Umwelt's approach enables a user to choose the set of representations that best suits their goals.
\section{Evaluation: User Study}

To evaluate Umwelt, we conducted remote studies with 5 expert BLV participants.
Each participant met with us for two 90-minute sessions over Zoom video calling with screenshare for a total of 3 hours per participant.
We split studies into two sessions to limit participant fatigue, and to give participants adequate time to become familiar enough with the system and its concepts to surface insights that reflect ordinary use conditions.
In the first session, participants used the \viewer{} to analyze an example dataset using multiple modalities.
In the second session, participants used the structured editor to choose from a set of example datasets and create their own multi-modal representations.
Participants were compensated \$250 for 3 hours.
The purpose of the evaluation was primarily exploratory, seeking to form an initial understanding of how screen reader users approach multi-modal representations and their specification.

Because Umwelt is a tool designed with expert users in mind, we made an intentional choice to recruit a smaller number of participants to spend more time going in-depth with each participant.
In qualitative research, the goal of selecting each qualitative case to examine is not to systematically answer descriptive questions about a population; it is to ``ask how or why questions about processes unknown before the start of the study'' \cite{small_how_2009}.
Consequently, the goal of recruitment in our study was not to create the largest, most representative sample of a population, but to draw on participants' lived experiences to reach a ``saturation'' of insights --- building our understanding to a point at which adding more participants stops giving us new or surprising information \cite{small_how_2009}.
We found that three hours per participant with five participants enabled us to reach saturation.

Recognizing that there is a history of exploitative relationships between researchers and marginalized research participants \cite{ymous_i_2020, liang_embracing_2021}, we reference our participants by name throughout the paper with their consent.
As scholars of citational justice in HCI note, a lack of intentional decision-making about who to acknowledge for their intellectual contributions can lead to the erasure of marginalized individuals' work and knowledge \cite{kumar_braving_2021}.
Our intention with this choice is to appropriately acknowledge and credit the expertise for which we recruited.
Following methodological recommendations to be specific about our target population and concept of expertise \cite{burns_who_2023}, our study conceives of expertise along two dimensions: screen reader experience, and data analysis experience.
We describe our participants' backgrounds in \autoref{tab:participants}.

\begin{table*}
  \caption{Participants' names, demographic information, and descriptions of their screen reader and data analysis experience. Participants are referenced by name with their consent.}
  \label{tab:participants}
  \renewcommand{\arraystretch}{1.2} %
  \begin{tabular*}{\textwidth}{@{\extracolsep{\fill}} p{0.1\textwidth} p{0.05\textwidth} p{0.25\textwidth} p{0.25\textwidth} p{0.25\textwidth}}
    \toprule
    Name & Age Bracket & Self-Description of Disability & Screen Reader Experience & Data Analysis Experience \\
    \midrule
    Ben Mustill-Rose & 20-35 & Totally blind, lost sight in early teens & Proficient with NVDA and sonification add-ons & Frequently analyzes data in Python as software engineer \\
    Ken Perry & 50+ & 100\% blind, lost sight in early 20s & Proficient with JAWS and other screen readers & Teaches Python and other programming languages, has written statistical software \\
    Dorene Cornwell & 50+ & Low vision / high partial, had detached retinas in mid-life & Proficient with JAWS screen reader + ZoomText for magnification & Masters-level courses in statistics and related fields \\
    Liam Erven & 35-50 & No usable vision, hearing impaired in right ear & Teaches students how to use assistive technology, including all major screen readers & Basic proficiency, uses spreadsheets \\
    Amy Bower & 50+ & Low partial vision, cannot see computer screen. Declining vision since mid-20s & Proficient in JAWS (self-taught) & Does research on oceanographic data. Uses Matlab for data analysis. \\
    \bottomrule
  \end{tabular*}
  \vspace{6pt}
\end{table*}

\subsection{Quantitative Results}
We designed two Likert surveys to separately evaluate the user experience of the \viewer{} and the editor. Participants responded on a scale of 1 to 5, where a higher number corresponds to an easier or more enjoyable experience.
We report participants' responses in \autoref{tab:quant-results}.
The median scores suggest that participants generally found both the \viewer{} and editor fairly easy to learn and enjoyable to interact with.
According to participants, the \viewer{} facilitates trend and pattern exploration in the data, and transitioning between modalities within the \viewer{} is straightforward. While participants rated the editor as slightly more difficult to learn, many also expressed interest in investing more time to learn because of its capabilities. When it comes to making edits, participants found the sonification settings easy to customize, but had a harder time predicting updates in the \viewer{} based on the changes made in the editor. In the qualitative analysis section, we will further contextualize participants’ ratings. 

\begin{table*}
  \caption{Rating scores for the \viewer{} and editor on a five-point Likert scale where {1} = Very Difficult (Very Unenjoyable) and {5} = Very Easy (Very Enjoyable). Median scores are shown in \textbf{bold}, averages in brackets [], standard deviations in parentheses ().}
  \label{tab:quant-results}
  \renewcommand{\arraystretch}{1.2} %
  \begin{tabular*}{\textwidth}{@{\extracolsep{\fill}} p{0.3\textwidth} p{0.15\textwidth} p{0.3\textwidth} p{0.15\textwidth}}
    \toprule
    Viewer & Score & Editor & Score \\
    \midrule
    How easy was it to learn to use the \viewer{}? & \textbf{4} [4] (0.71) & How easy was it to learn to use the editor? & \textbf{3} [3.4] (0.55) \\
    After understanding how the \viewer{} works, how enjoyable was it to interact with the data? & \textbf{5} [4.8] (0.45) & After understanding how the editor works, how enjoyable was it to edit the data representation? & \textbf{4} [3.8] (0.45) \\
    After understanding how the \viewer{} works, how easy was it to switch between descriptions and sonifications on-demand? & \textbf{4} [4.2] (0.45) & If you had a change you wanted to make to the data representation, how easy is it to understand how to make that change using the editor? & \textbf{4} [3.8] (0.84) \\
    How easy was it to be able to customize the sonification settings, including playback mode, audio axis ticks, and playback rate? & \textbf{5} [4.6] (0.55) & After understanding how the editor works, how easy was it to predict how changes in the editor would affect the \viewer{}? & \textbf{3} [3.6] (0.89) \\
    After understanding how the \viewer{} works, how easy was it to understand trends and patterns in the data? & \textbf{4} [4.4] (0.55) & After understanding how the editor works, how easy was it to check the result of your edits in the \viewer{}? & \textbf{4} [4] (0.71) \\
    \bottomrule
  \end{tabular*}
  \vspace{6pt}
\end{table*}

\subsection{Qualitative Results: Multi-Modal Viewer}
\label{sec:qualitative-results}

\subsubsection{Modalities have complementary affordances.}
Participants found it useful to have multiple modalities available for many reasons, including increased optionality, modality-specific affordances, complementary uses leading to better understanding, and toggling between overview and detail.

\textbf{Multiple representations as options to accommodate varying needs.}
Offering multiple modalities can help avoid cognitive or sensory overload.
Minimizing cognitive load is a foundational principle in HCI; however, as research on accessible data analysis has shown, cognitive load can pose amplified challenges when it intersects with various disabilities \cite{doug_schepers_designing_2022}.
As Erven noted, using only textual or tabular representations can result in ``number fatigue'' where the numbers ``stop meaning anything.''
This fatigue can be compounded for people with disabilities related to attention management or memory.
Having the option to switch from textual representations to sonification can potentially help provide more usable alternatives.

For users who may need to commit additional effort to use certain representations, having alternatives can also help manage sensory fatigue.
Perry, who works with low-vision colleagues, suggested that they might like to ``rest [their] eyes [while] flipping through the data.''
In these situations, being able to switch to a different representation can better accommodate an individual's needs by balancing their sensory load.

\textbf{Complementary modalities enable better understanding via overview and detail.}
Just like sighted visualization users, studies \cite{zong_rich_2022, sharif_understanding_2021} have shown that BLV users follow the information-seeking heuristic of ``overview first, zoom and filter, and details on demand'' \cite{shneiderman_eyes_2003}.
As Bower noted, ``when people look at a graph, they look at the big picture first and then they start scrutinizing it.''
Participants found that sonification and textual description complement each other by effectively conveying overview and detail, respectively.
Mustill-Rose noted that the textual description gave him the min, max, and average values, which are ``hard if not impossible to get from sonification.''
On the other hand, Perry enjoyed the ability to sonify ``trend lines in the data without having to go point by point.''
Since modalities afford different kinds of information-seeking operations, participants sought to choose the right modality for the task at hand.
Switching between representations also helped participants adjust their initial assumptions about the data.
For instance, Mustill-Rose listened to the sonification first and initially hypothesized that the stocks dataset contained only one data point per year.
Then, he noticed that this was not the case when he explored the textual representation.
He reflected that ``the lesson there is to not consider just one modality.''

\subsubsection{Synchronized query predicates help users share context between modalities.}
Participants valued the ability to maintain a shared query predicate while switching between modalities, which crucially helped them think of the modalities as different ways of looking into the same underlying data.
Mustill-Rose described the synchronization across modalities as an ``enabler'' in the sense that ``it's decreasing the time that it's taking me to get the data [from] the [time] period that I need'' before he then ``switch[es] to something else to look at it in a different way.''
Because the system maintained his interactive context as he switched representations, he was able to stay in the flow of his ongoing analysis.
This echoes prior findings that delays caused by interactive latency during data analysis can ``[disrupt] fluent interaction'' and cause people to lose their train of thought during exploratory analysis \cite{liu_immens_2013}.

\subsubsection{Customization supports differences in task and experience.}
Research has shown that customizable textual descriptions support users who have different preferences or tasks, allowing them to control the information they receive and how it's presented \cite{jones_customization_2023}.
This was reiterated by Perry, Cornwell, and Erven, who encountered situations where they wanted to adjust the presence, verbosity, and ordering of information in text.
We also found this customizability idea applicable beyond textual description, particularly for Jones et al.'s wayfinding and consuming affordances \cite{jones_customization_2023}.

\textbf{Wayfinding.}
The audio axis ticks feature supports wayfinding by helping users understand their progress through a temporally proceeding sonification.
However, it trades off efficiency, and becomes less necessary over time as users get more familiar with the data.
Mustill-Rose found himself wanting to disable the axis ticks after listening to a few sonifications.
He said, ``at first it was useful [...] but now that I know what I'm looking at, I feel like the [audio axis ticks] has proved its value. And now I don't need it anymore.''
However, once he selected a different subset of the data, he realized that it was ``now useful again because I haven't explored this section.''
His need for this feature was situational throughout his analysis, depending on whether he was focusing on learning the layout of the data or ``understanding and honing in on'' the data.
As a result, the ability to enable or disable the axis ticks was important to offer as a customization.

\textbf{Consuming.}
Another important customization was the sonification's playback speed.
As Cornwell noted, preferred screen reader reading speed varies widely among BLV individuals.
For sonification, participants considered their base preference as well as their familiarity and task-specific needs.
Perry and Bower both noted that their preferred playback speed was situational.
Perry noted that he ``would get used to it faster, but [he] would start slower because it gives more time to listen to each point.''
Slower speeds were better when initially learning about the data, and he would speed up as he became more familiar.

\subsubsection{Multi-modal representations facilitate communication between people who rely on different senses.}
In a multi-modal system, participants who were not primarily using the visualization still valued the presence of synchronized visual highlighting and references to visual concepts in the description.
As BLV professionals who work with sighted colleagues, participants frequently encounter situations where they need to establish common ground with others who primarily use different senses.

\textbf{Contributing confidently to data-driven discussions.}
As a software engineer who works with only sighted colleagues, Mustill-Rose stressed that an important goal of data analysis is to have enough information to ``participate meaningfully in a discussion.''
At minimum, he said, he wanted to be in Zoom meetings and ``not seem clueless,'' because as the only blind person on a team, consistently being the only person who can't comment on a topic can compound with unconscious bias to affect promotions and work opportunities.
Erven echoed this sentiment, saying that ``the most important thing is independence.''
Visual modalities are only helpful when ``it's not something you need to rely on to do your work,'' forcing BLV users to rely sighted help.
Instead, as previous work has also argued \cite{zong_rich_2022}, accessible representations should promote user agency for self-guided analysis --- and for BLV people not only to participate in, but also create and lead data-driven discussions.

\textbf{Presenting to mixed audiences.}
In her job, Cornwell frequently makes presentations to majority sighted audiences.
As a result, ``visual charts are always useful'' to her.
She explained, ``if I'm needing to talk about [the data], I can just say, look at the red line and the people who are really visual --- that's an important source of interactivity for them.''
Additionally, synchronization between modalities plays a helpful role in presentation.
For example, Cornwell imagined a hypothetical situation where she played a sonification while presenting, and sighted audience members followed along on the visualization.
In this situation, having multiple modalities would make the presentation more accessible and also help communicate the data more effectively.

\textbf{Collaborating across different levels of vision.}
Many participants frequently collaborate directly with others with different levels of vision.
Cornwell mentioned working with someone who was totally blind, and thought that ``sonification on a screen share'' would be extremely valuable for communicating about data.
Similarly, Bower felt that the visual highlighting of her selection in the textual structure and sonification would help a collaborator ``get on the same page'' and help them ``know where [she's] looking.''
She drew an analogy to pointing at something on a visual chart, as a way of directing a collaborator's attention.

\subsection{Qualitative Results: Structured Editor}

\subsubsection{Users want, but lack, interfaces for creating data representations.}
Participants have existing strategies for working with data that primarily involve spreadsheets and scripting.
Erven, Perry, Cornwell, and Bower reported using Excel or Google Sheets; Mustill-Rose, Perry, and Bower reported writing their own scripts in various tools, including python, octave, and matlab.
However, there was consensus that these existing workflows are insufficient.
Cornwell put it succinctly when describing raw data: ``no one wants to read that stuff.''
But with the exception of Bower (who has used Highcharts Sonification Studio and SAS Graphics Accelerator), no participants could think of available tools for creating their own representations without having to write code.

End-user tools are important because they lower the technical barrier for creating representations. When comparing the Umwelt editor to writing code, Mustill-Rose said that ``there's less cognitive pressure using a UI than if I was having to write code to do it.''
However, sometimes tools can overly complicate the process of making a simple representation.
Bower said, ``I don't care about instruments and timbre and all that, I just want access to a time series.''
Because of high up-front specification cost, some tools are too difficult to use for even simple cases.
Nonetheless, Bower is interested in trying new tools for creating data representations, saying, ``I'm kind of desperate for anything'' that's usable and accessible.

\subsubsection{Structured editing with default specifications reduces semantic and articulatory distance.}
When participants decide to create representations, they face challenges to do with semantic and articulatory distance \cite{hutchins_direct_1985}. 
In HCI theory, semantic distance is the distance between a user's intentions and how these intentions are translated into the concepts provided by a user interface.
Similarly, articulatory distance is the relationship between an interface's concepts and the set of physical actions a user has to take to express something in terms of those concepts.

\textbf{Semantic distance.}
When Perry approached analyzing the penguin dataset, he initially said, ``I want to compare beak length, body mass, and sex altogether --- I want to see the graph for all three of these together.''
Though Perry knew that he wanted to specify visual encodings that would be reflected in the textual hierarchy, he did not immediately know what those encodings were.
This was a problem of semantic distance, because he needed to map his goal onto the concepts provided by the user interface.
Luckily, the heuristics generated a default specification for that set of fields that matched his expectations.
As a result, he was still able to create the chart despite lack of familiarity with visualization concepts.
However, when the system was not able to generate a default specification for Cornwell, she remarked that it was hard to figure out ``which functions apply, like figuring out if I wanted it grouped by island or species.''
Even though she had goals in mind, it was difficult for her to translate those goals into specific encodings and field transformations.
This suggests a need for future work on bridging semantic distance --- for instance, by designing high-level abstractions that adhere closer to users' abstract goals, reducing the amount of translation work.

\textbf{Articulatory distance.}
Using the editor, Mustill-Rose remarked that ``if I were writing code, I'd need to think about what the end result was and what code I needed to write to achieve it at the same time.''
Rather than having to remember the names of functions and expressions in a textual language, Mustill-Rose was able to use commonplace HTML input elements that express atomic edits to a specification as simple button clicks or dropdown selections.
However, the editor also has limitations when it comes to articulatory distance.
Cornwell, who had created a chart that was not a default specification, noted that a main challenge was that ``when you’re looking at the fields, you have to add encodings for everything you want.''
When specifying multi-modal representations, there can be a lot of repetitive operations to create three outputs that are conceptually similar.

\subsubsection{Users think in both field-oriented and encoding-oriented terms.}
Throughout the specification process, we observed that participants moved between field-oriented and encoding-oriented ways of thinking.
For most participants, the tendency was to begin by identifying a set of fields they were interested in.
For instance, Erven commented that it felt natural to begin by ``choosing the fields you want,'' since you ``might not want all that data.''
When default specifications matched their expected representation, or when they only required minor edits, participants were generally content with the result that they achieved through field-oriented specification.

However, when more manual editing was required to achieve the desired output, we found that participants shifted more toward encoding-oriented specification as they envisioned specific output representations.
Cornwell initially stated her goal by saying, ``I want to know what species are on which island and then I want a sex distribution.''
At this point in his process, she had not committed to any encoding properties or specific modalities, but was envisioning the semantics and structure of the data in terms of relationships between fields.
After selecting the relevant fields, she began to add encodings, and then became somewhat stuck.
When prompted to reiterate her goal, she said that she wanted to create a ``bar chart with island on the x-axis and count for species for the y-axis.''
At this point further into the process, she had imagined a specific visual representation, which she was attempting to decompose into encodings and then map onto editor operations.

Interestingly, Bower\,---\,who is familiar with both visualization and sonification\,---\,had a mental model that blurred the dichotomy between field- and encoding-oriented specification.
She initially approached the Seattle weather dataset by selecting \texttt{date} and \texttt{temp\_max}.
When she tabbed down the editor to read the default specification, she noticed that Umwelt had assigned \texttt{y} and \texttt{pitch} encodings to \texttt{temp\_max}.
Based on her extensive previous experience with data visualization and sonification, she remarked, ``I merge those in my head -- I think of those as the same thing.''
This suggests that even when thinking in encoding-oriented terms, Bower was reasoning about the data's underlying key-value semantics.

\section{Discussion and Future Work}

We contribute Umwelt, an accessible authoring environment designed to de-center the visual modality in data analysis. Umwelt allows users to specify data representations, including visualization, structured textual description, and sonification, using a shared abstract data model. Unlike existing tools, Umwelt does not rely on an existing visual specification, affording users more flexibility in prototyping multi-modal representations. The editor's state is reflected in independent visual, textual, and sonification views linked through shared interactions, encouraging complementary use of multiple modalities.
In this section, we discuss potential directions for future work surfaced by Umwelt.

\subsection{Designing Natively Non-Visual Data Representations}

Differences in a representation's modality affect how information is presented to a user, and the operations the user needs to perform to access the information. For example, a screen reader must ``explicitly linearize reading a visualization'' in order to narrate elements one at a time \cite{zong_rich_2022} --- in contrast to how visual perception enables a user to move around parts of a visualization relatively freely. Similarly, researchers have compared tactile perception to ``reading a map through a small tube'' \cite{zong_rich_2022, hasty_guidelines_2011}.
An implication of these modality differences, as participants in our study found, is that users find some modalities inherently more suited to certain tasks than others.
Further, it suggests that due to medium-specificity, it is not always possible to directly translate a data representation from one modality to another while maintaining 1:1 support for the same set of tasks.

However, existing systems for authoring non-visual representations largely attempt to directly translate source visualizations into standalone non-visual replacements.
For example, while Highcharts Sonification Studio \cite{cantrell_highcharts_2021} successfully translates single-series line charts into equivalent sonifications, this approach breaks down for scatterplots.
This is because the way a sighted user reads a scatterplot has no unambiguous analogue in the medium of sonification, which imposes a linearized traversal order over the data.
In contrast, Umwelt's default specification heuristics pair scatterplots with sonifications that diverge from the visualization by adding additional binning and aggregation, in order to prioritize conveying the data's 2d distribution.

In the context of multimodal representations, Umwelt advances the idea that a representation should prioritize fit with its modality's affordances over fidelity to the visual representation.
This conceptual orientation has implications for the design of future non-visual representations.
For example, current approaches to tactile charts largely focus on converting visual channels to tactile ones while otherwise faithfully recreating the visualization \cite{engel_svgplott_2019}.
Instead, future work could explore tactile-first designs that make more intentional use of the processual, part to whole \cite{hasty_teaching_2017} nature of tactile perception.

\subsection{Interdependence and Relational Dimensions of Access}

In designing Umwelt, we advocate for a conceptual shift in the field of accessible data visualization --- focusing not only on making existing visualization accessible to BLV readers, but also on empowering BLV data analysts to independently produce their own representations and conduct self-guided data exploration.
Because existing approaches that center the visual modality can sometimes create barriers or reinforce BLV users' dependence on sighted assistance, we believe a focus on independence to be an important step forward.
However, in addition to independence, disability scholars have advanced \textit{interdependence} as a complementary conceptual frame \cite{bennett_interdependence_2018}.
An interdependence frame acknowledges that all people constantly depend on others, and so a focus on relationships is necessary to understand how access is socially created in practice.

Our initial evaluation of Umwelt surfaced ideas that suggest the need for interdependence (alongside independence) as a lens for design.
For instance, participants felt that building common ground between mixed ability colleagues in workplace settings was important to their career advancement, highlighting the fact that BLV people's access needs are embedded in a social and relational context.
Future work can motivate and inspire system design based on not only how BLV users can get immediate access to information in data, but also how they hope to use that information to participate in broader social processes.

\begin{acks}
Thank you to Josh Pollock, Amy Fox, and Shuli Jones. This work was supported by NSF \#1942659 and \#1900991, and the MIT Morningside Academy for Design Fellowship.
\end{acks}

\bibliographystyle{ACM-Reference-Format}
\bibliography{umwelt}

\end{document}